\documentclass[floats,prb,twocolumn,amsmath,showpacs]{revtex4}

\usepackage{epsfig}
\usepackage{colordvi}
\usepackage{graphicx}
\def\k{{\bf k}}

\def\R{{\bf R}}
\def\r{{\bf r}}
\def\b{{\bf b}}
\def\q{{\bf q}}
\def\pw{^{({\rm W})}}
\def\ph{^{({\rm H})}}
\def\la{\langle\kern-2.0pt\langle}
\def\ra{\rangle\kern-2.0pt\rangle}
\def\vt{\vert\kern-1.0pt\vert}
\def\D{{D}\ph}
\begin{document}
\def\dvm#1{\marginpar{\small DV: #1}}
\def\xwm#1{\marginpar{\small XW: #1}}
\def\jry#1{\marginpar{\small JY: #1}}
\def\ivo#1{\marginpar{\small IS: #1}}

\title{Ab initio calculation of the anomalous Hall
conductivity by Wannier interpolation}

\author{Xinjie Wang,$^1$ Jonathan R. Yates,$^{2,3}$ Ivo Souza,$^{2,3}$ and
David Vanderbilt$^1$}
\affiliation{$^1$Department of Physics and Astronomy, Rutgers University,
        Piscataway, NJ 08854-8019\\
        $^2$Department of Physics, University of California, Berkeley, CA 94720\\
        $^3$Materials Science Division, Lawrence Berkeley National Laboratory, Berkeley,
        CA 94720}
\date{\today}
\begin{abstract}
The intrinsic anomalous Hall effect in ferromagnets depends on subtle 
spin-orbit-induced effects in the electronic structure, and 
recent {\it ab-initio} studies found that it was necessary to 
sample the Brillouin zone at millions of $k$-points to converge the
calculation.
We present an efficient first-principles approach for computing the
anomalous Hall conductivity.
We start out by performing
a conventional electronic-structure calculation including spin-orbit coupling
on a
uniform and relatively coarse
$k$-point mesh.  From the resulting Bloch states,
maximally-localized Wannier functions are constructed
which reproduce the {\it ab-initio} states up to the Fermi level.
The Hamiltonian and position-operator matrix elements, needed to
represent the
energy bands and Berry curvatures,
are then set up between the Wannier orbitals.
This completes the first stage of the calculation, whereby
the low-energy
{\it ab-initio} problem is transformed into an effective tight-binding form.
The second stage only involves Fourier transforms and
unitary transformations of the small matrices set up  in the first stage.
With these inexpensive operations, the quantities of interest are
interpolated onto a dense $k$-point mesh and used to evaluate the
anomalous Hall conductivity as a Brillouin zone integral.
The present scheme, which also avoids the
cumbersome summation over all unoccupied states in the Kubo formula,
is applied to bcc Fe,
giving excellent agreement with conventional, less efficient
first-principles calculations.
Remarkably, we find that more than 99\% of the effect can be recovered by 
keeping a set of terms depending only on the Hamiltonian matrix elements,
not on matrix elements of the position operator.
\end{abstract}


\maketitle

\vskip2pc
\marginparwidth 3.1in
\marginparsep 0.5in

\section{Introduction}

The Hall resistivity of a ferromagnet depends not only on the
magnetic induction, but also on the magnetization; the latter dependence
is known as the anomalous Hall effect (AHE).\cite{hurd72} 
The AHE is used for investigating surface magnetism, and its potential
for investigating nanoscale magnetism, as well as for magnetic sensors and 
memory devices applications, is being considered.\cite{gerber02} 
Theoretical investigations of the AHE have undergone a revival in
recent years, and have also lead to the proposal for a spin counterpart,
the spin Hall effect, which has subsequently been realized experimentally.

The first theoretical model of the AHE was put forth by Karplus and
Luttinger,\cite{karplus54} who showed that it can arise in a perfect crystal
as a result of the spin-orbit interaction of polarized conduction
electrons. 
Later, two alternative mechanisms, skew scattering\cite{smit58}
and side jump scattering,\cite{berger70}
were proposed by Smit and Berger respectively.
In skew scattering the spin-orbit interaction gives rise to an
asymmetric scattering cross section even if the defect potential
is symmetric, and in side-jump scattering the spin-orbit coupling causes the
scattered electron to acquire an extra transverse translation after
the scattering event. These two mechanisms
involve scattering from impurities or phonons, while the 
Karplus-Luttinger mechanism is a scattering-free bandstructure
effect.  For reasons related to the absence of an intuitive physical
picture and the lack of reliable quantitative estimates based on
bandstructure calculations, the Karplus-Luttinger theory was strongly 
disputed in the early literature.

In recent years, new insights into the Karplus-Luttinger mechanism have
been obtained by
several authors,\cite{chang96,sundaram99,onoda02,jungwirth02,haldane04} who
reexamined it in the modern language of Berry's phases.
The term ${\boldsymbol\Omega}_n(\k)$ in the equations below
was recognized as the Berry curvature of the Bloch states
in reciprocal space, a quantity which had previously appeared in the
theory of 
the integer quantum Hall effect,\cite{thouless82} and also in the
Berry-phase theory of polarization.\cite{ksv93}
The anomalous Hall conductivity (AHC)
is simply given as the Brillouin zone (BZ) integral of the Berry curvature weighted
by the occupation factor of each state,
\begin{eqnarray}
                \sigma_{xy}=\frac{-e^2}{(2\pi)^2h}\sum_{n}\int_{\rm BZ}\,
d\k \, f_n(\k)\,\Omega_{n,z}(\k) \;.
\label{eq:sigma}
\end{eqnarray}
While this can be derived in several ways, it is perhaps most
intuitively understood from the semiclassical point of view, in which
the group velocity of an electron wavepacket in band $n$ 
is\cite{adams59,sundaram99}
\begin{equation}
 {\dot \r}=\frac{1}{\hbar}\frac{\partial {\cal E}_{n\k}}{\partial \k}-
        {\dot \k}\times {{\bf \Omega}_n(\k)}\;.
\label{eq:rdot}
\end{equation}
The second term, often overlooked in elementary textbook derivations,
is known as the ``anomalous velocity.''  The expression for
the current density $\bf J$ then acquires a new term
$ef_n(\k)\,{\dot \k}\times {\bf \Omega}_n(\k)$ which, with
${\dot \k}=-e{\bf E}/\hbar$, leads to Eq.~(\ref{eq:sigma}).

Recently, first-principles calculations of Eq.~(\ref{eq:sigma}) were
carried out for the ferromagnetic perovskite SrRuO$_3$ by
Fang {\it et al.},\cite{fang03} and for a transition metal, bcc Fe, by Yao
 {\it et al.}\cite{yao04} 
In both cases the calculated values compared well with experimental data,
lending credibility to the intrinsic mechanism.
The most striking feature of these calculations is
the strong and rapid variation of the Berry curvature 
in $k$-space. In particular, there are sharp peaks and valleys at places
where two energy bands are split by the spin-orbit coupling across the Fermi 
level. In order to converge the integral, the Berry curvature has to be
evaluated over millions of $k$-points in the Brillouin zone. In the previous
work this was done via a Kubo formula involving a large number of unoccupied
states; the computational cost was very high, even for bcc Fe, with only one
atom in the unit cell.

In this paper, we present an efficient method for computing the AHC. 
Unlike the conventional approach, it does not require carrying out a full 
{\it ab-initio} calculation for every $k$-point where the Berry
curvature needs to be evaluated. The actual
{\it ab-initio} calculation is performed on a much coarser $k$-point grid. 
By a post-processing step, the resulting Bloch states below
and immediately above the Fermi level are
then mapped onto well-localized Wannier-functions.  
In this
representation it is then possible to interpolate the Berry curvature onto any
desired $k$-point with very little computational effort and essentially
no loss of accuracy.

The paper is organized as follows. In Sec.~\ref{sec:background} we
introduce the basic definitions and describe the Kubo-formula approach
used in previous calculations of the intrinsic AHC.
In Sec.~\ref{sec:fdbc} our new Wannier-based approach is described.
The details of the band-structure calculation and Wannier-function
construction are described in
Sec.~\ref{sec:cd}, followed by an application of the method to
bcc Fe in Sec~\ref{sec:results}.
Finally, Sec.~\ref{sec:conclusion} contains a brief
summary and discussion.

\section{Definitions and background}
\label{sec:background}

The key ingredient in the theory of the intrinsic anomalous Hall effect is
the Berry curvature ${\bf \Omega}_n({\bf k})$, defined as
\begin{eqnarray}
        {\bf \Omega}_n({\bf k})=\boldsymbol\nabla\times {\bf A}_n(\k) \;,
        \label{eq:bc}
\end{eqnarray}
where ${\bf A}_{n}$ is the Berry connection,
\begin{eqnarray}
        {\bf A}_n(\k)=i\langle u_{n\k}|\boldsymbol\nabla_\k|u_{n\k} \rangle\;.
        \label{eq:berrypot}
\end{eqnarray}
The Berry curvature can be written in an
equivalent but more explicit form:
  \begin{equation}
  {\Omega}_{n,\gamma}({\bf k})=\epsilon_{\alpha \beta \gamma}
        \, \Omega_{n,\alpha \beta}(\k) \;,
  \end{equation}
  \begin{equation}
        \Omega_{n,\alpha \beta}(\k) =
        -2\,{\rm Im}\,\Big\langle \frac{\partial u_{n\k}}
        {\partial k_\alpha}
        \Big|\frac{\partial u_{n\k}}{\partial k_\beta} \Big\rangle \;,
  \label{eq:bcurv}
  \end{equation}
where the Greek letters indicate Cartesian coordinates,
$\epsilon_{\alpha \beta \gamma}$ is Levi-Civita tensor and
$u_{n\k}$ are the cell-periodic Bloch functions. The second-rank
Berry curvature tensor $\Omega_{n,\alpha \beta}(\k)$ is introduced
for later use.  The integral of the Berry
curvature over a surface bounded by a closed path $C$ in $k$-space
is the Berry phase of that path.\cite{berry84}

With this notation we rewrite the quantity we wish to evaluate,
Eq.~(\ref{eq:sigma}), as
\begin{eqnarray}
                \sigma_{\alpha\beta}=\frac{-e^2}{(2\pi)^2h}\int_{\rm BZ}\,
d\k\, \Omega_{\alpha\beta}(\k) \;,
\label{eq:sigma_b}
\end{eqnarray}
where we have introduced the {\it total} Berry curvature
\begin{equation}
\label{eq:omega_tot}
\Omega_{\alpha\beta}(\k)=\sum_{n}\,f_n(\k)\,\Omega_{n,\alpha\beta}(\k).
\end{equation}
Direct evaluation of Eq.~(\ref{eq:bcurv})
poses a number of practical difficulties related to the presence of
$k$-derivatives of Bloch states, as will be discussed in the next section.
In previous work\cite{fang03,yao04} these were circumvented by recasting
Eq.~(\ref{eq:bcurv}) as a Kubo formula,\cite{thouless82} where
the $k$-derivatives are replaced by sums over states:
\begin{eqnarray}
        \Omega_{n,\alpha\beta}({\k})=-2{\rm Im}\sum_{m \neq n}
   \frac{v_{nm,\alpha}(\k)\,v_{mn,\beta}(\k)}
        {(\omega_{m}(\k)-\omega_{n}(\k))^2}\;,
         \label{eq:kubo}
\end{eqnarray}
where $\omega_{n}(\k)={\cal E}_{n\k}/\hbar$ and the 
matrix elements of the Cartesian velocity operators
$\hat v_\alpha=(i/\hbar)[\hat H,\hat r_\alpha]$ are given by\cite{blount62}
\begin{equation}
v_{nm,\alpha}(\k)=\langle \psi_{n\k}|\hat v_\alpha|\psi_{m\k}\rangle=
\frac{1}{\hbar}\,
\Big\langle u_{n\k}\Big|\frac{\partial \hat H(\k)}{\partial k_\alpha}\Big|
u_{m\k}\Big\rangle\;,
\label{eq:vel}
\end{equation}
where $\hat H(\k)=e^{-i\k\cdot\hat\r}\hat{H}e^{i\k\cdot\hat\r}$.
The merit of Eq.~(\ref{eq:kubo}) lies in its practical implementation on a
finite $k$-grid using only the wave functions at a single $k$-point.
As is usually the case for such linear-response
formulas, sums over pairs of occupied states can be avoided in the
$T=0$ version of the formula (\ref{eq:omega_tot}) for the total
Berry curvature,
\begin{eqnarray}
        \Omega_{\alpha\beta}({\k})=-2{\rm Im}\sum_{v}\sum_{c}
   \frac{v_{vc,\alpha}(\k)\,v_{cv,\beta}(\k)}
        {(\omega_{c}(\k)-\omega_{v}(\k))^2} \;,
         \label{eq:kubotot}
\end{eqnarray}
where $v$ and $c$ subscripts denote valence (occupied) and conduction 
(unoccupied) bands, respectively.  However, the evaluation of this formula
requires the cumbersome summation over unoccupied states.  Even if
practical calculations truncate the summation to some extent, the
computation could be time-consuming. Moreover, the time required to
calculate the matrix elements of the velocity operator in Eq.~(\ref{eq:kubo})
or (\ref{eq:kubotot}) is not negligible.

\section{Evaluation of the Berry curvature by Wannier interpolation}
\label{sec:fdbc}

In view of the above-mentioned drawbacks of the Kubo formula for
practical calculations, it would be highly desirable to have a
numerical scheme based on the
the ``geometric formula'' (\ref{eq:bcurv}), in terms of the
occupied states only. The difficulties in implementing that
formula arise form the $k$-derivatives therein.
Since in practice one always replaces the Brillouin zone integration by a
discrete summation, an obvious approach would be to use a finite-difference
representation of the derivatives on the $k$-point grid. However, this requires
some care: a straightforward discretization will yield results which
depend on the choice of phases of the Bloch states (i.e., the choice of
gauge), even though Eq.~(\ref{eq:bcurv}) is in principle gauge-invariant.
The problem becomes more acute in the presence of band crossings and
avoided crossings, because
then it is not clear which two states at neighboring grid points should
be taken as ``partners'' in a finite-differences expression. 
(Moreover, since the system is a metal, at $T=0$ the occupation can be
different at neighboring $k$-points.) Successful numerical strategies 
for dealing with problems of this nature have been developed in the 
context of the Berry-phase theory of polarization of insulators, and 
a workable  finite-difference scheme which combines those ideas 
with Wannier interpolation is sketched in Appendix~B.

We present here a different, more powerful strategy that also relies 
on a Wannier representation of the low-energy electronic structure.  
We will show that it is possible to express the needed derivatives 
analytically in terms of the Wannier functions, so that no finite-difference 
evaluation of a derivative is needed in principle. The use of Wannier
functions
allows us to achieve this while still avoiding the summation over all
empty states which appears in the Kubo formula as a result of applying
conventional $k\cdot p$ perturbation theory.

\subsection{Wannier representation}
\label{sec:wr}

\begin{figure}
\begin{center}
\epsfig{file=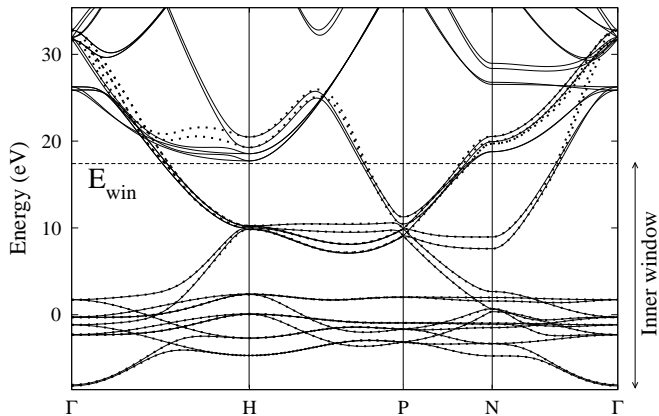,width=3.4in}
\end{center}
\caption{Band structure of bcc Fe with spin-orbit coupling
included. Solid lines: original band structure
from a conventional first-principles calculation.
Dotted lines: Wannier-interpolated band
structure. The zero of energy is the Fermi
level.}
\label{fig:band}
\end{figure}

We begin by using the approach of Souza, Marzari, and Vanderbilt
\cite{souza01} to construct a set of Wannier functions (WFs)
for the metallic system
of interest.  For insulators, one
normally considers a set of WFs that span precisely the space of
occupied Bloch states.  Here, since we have a metallic system and
we want to have well-localized WFs, we choose a number of WFs
larger than the number $N_\k$ of occupied states at any $\k$, and 
only insist that the
space spanned by the WFs should include, as a subset, the space
of the occupied states, plus the first few 
empty states. Thus, these partially-occupied
WFs will serve here as a kind
of ``exact tight-binding basis'' that can be used
as a compact representation of
the low-energy electronic structure of the metal.

This is illustrated in Fig.~\ref{fig:band}, where the bandstructure
of bcc Fe is shown.  The details of the calculations will be presented
later in Sec.~\ref{sec:cd}.
The solid lines show the full {\it ab-initio} bandstructure, while
the dashed lines show the bands obtained within the Wannier
representation using $M=18$ WFs per cell (9 of each spin;
see Sec.~\ref{sec:maxloc}).
In the method of Ref.~\onlinecite{souza01},
one specifies an energy $E_{\rm win}$ lying somewhat
above the Fermi energy $E_{\rm f}$, and insists on finding a set
of WFs spanning all the states in an energy window up to $E_{\rm win}$.
In the calculation of Fig.~\ref{fig:band} we chose $E_{\rm win}\simeq
18$\,eV, and it is evident that there is an essentially perfect match
between the fully {\it ab-initio} and the Wannier-represented bands up to,
but not above, $E_{\rm win}$.

More generally, we shall assume that we have $M$ WFs per unit cell
(where $M\ge N_\k$ everywhere in the BZ)
such that the Bloch-like functions given by the phased sum of the
Wannier orbitals,
\begin{equation}
\label{eq:blochW}
|u_{n\k}\pw\rangle = \sum_\R e^{-i\k\cdot(\hat\r-\R)}\,|\R n\rangle
\end{equation}
($n=1,...,M$), span the actual Bloch eigenstates $|u_{n\k}\rangle$
of interest
($n=1,...,N_k$) at each $\k$.  It follows that, if we construct the
$M\times M$ Hamiltonian matrix
\begin{equation}
\label{eq:hamW}
H_{nm}\pw(\k)=\langle u_{n\k}\pw | \hat H(\k) | u_{m\k}\pw \rangle
\end{equation}
and diagonalize it by finding an $M\times M$ unitary rotation matrix
$U(\k)$ such that
\begin{equation}
U^\dagger(\k) H\pw(\k) U(\k) = H\ph(\k)
\label{eq:Htrans}
\end{equation}
where $H\ph_{nm}(\k)={\cal E}\ph_{n\k}\delta_{nm}$, then ${\cal E}\ph_{n\k}$
will be identical to the true ${\cal E}_{n\k}$ for all occupied
bands.  Also, the corresponding Bloch states
\begin{equation}
|u_{n\k}\ph\rangle = \sum_m |u_{m\k}\pw\rangle U_{mn}(\k)
\label{eq:twist}
\end{equation}
will also be identical to the true eigenstates $|u_{n\k}\rangle$ for ${\cal E}\le
E_{\rm f}$.  (In the scheme of Ref.~\onlinecite{souza01}, these
properties will actually hold for energies up to $E_{\rm win}$.)
However, the band energies and Bloch states will {\it not} generally
match the true ones at higher energies, as shown in Fig.~\ref{fig:band}.
We thus use the superscript `H' to distinguish the projected band
energies ${\cal E}\ph_{n\k}$ and eigenvectors $|u_{n\k}\ph\rangle$ from
the true ones ${\cal E}_{n\k}$ and $|u_{n\k}\rangle$, keeping in mind
that this distinction is only significant in the higher-energy unoccupied
region (${\cal E}>E_{\rm win}$) of the projected bandstructure.

The unitary rotation of states expressed by the matrix $U(\k)$ is
often referred to as a ``gauge transformation,'' and we shall adopt
this terminology here.  We shall refer to the
Wannier-derived Bloch-like states $|u_{n\k}\pw \rangle$ as belonging
to the Wannier (W) gauge, while the eigenstates $|u_{n\k}\ph \rangle$ of the
projected bandstructure are said to belong to the Hamiltonian (H)
gauge.

Quantities such as the Berry connection ${\bf A}_n(\k)$ of
Eq.~(\ref{eq:berrypot}) and the Berry curvature
$\Omega_{n,\alpha\beta}(\k)$ of Eq.~(\ref{eq:bc}) clearly depend
upon the gauge in which they are expressed.
The quantity that we wish to calculate,
Eq.~(\ref{eq:omega_tot}), is most naturally
expressed in the Hamiltonian gauge, where it takes the form
\begin{equation}
  \Omega_{\alpha\beta}(\k)=\sum_{n=1}^{M} f_n(\k) \,
  \Omega_{n,\alpha\beta}\ph(\k)
\;.
\label{eq:omsum}
\end{equation}
Here $\Omega_{n,\alpha\beta}\ph(\k)$ is given by Eq.~(\ref{eq:bcurv})
with $|u_{n\k}\rangle\rightarrow|u_{n\k}\ph\rangle$.  It is permissible
to make this substitution because the projected bandstructure matches the
true one for all occupied states.
In practice one may take for the occupation factor $f_n(\k)
=\theta(E_{\rm f}-{\cal E}_{n\k})$ or introduce a small thermal
smearing as desired.

Our strategy now is to see how
the right-hand side of Eq.~(\ref{eq:omsum}) can be obtained by starting
with quantities that are defined and computed first in the Wannier
gauge and then transformed into the Hamiltonian gauge.
The resulting scheme can be viewed as a generalized Slater-Koster 
interpolation, which takes advantage of the smoothness in $k$-space of the
Wannier-gauge objects, a direct consequence of the short range of the Wannier
orbitals in real space.

\subsection{Gauge transformations}

Because the gauge transformation of Eq.~(\ref{eq:twist}) involves a
unitary rotation among several bands, it is useful to introduce
generalizations of the quantities in Eqs.~(\ref{eq:bc}-\ref{eq:berrypot})
having two band indices instead of one. Thus, we define
\begin{equation}
A_{nm,\alpha}(\k)=i\langle u_{n}|\partial_\alpha u_{m}\rangle
\label{eq:Awg}
\end{equation}
and
\begin{equation}
\Omega_{nm,\alpha\beta}(\k)=
  i\langle \partial_\alpha u_{n}|\partial_\beta  u_{m}\rangle
 -i\langle \partial_\beta  u_{n}|\partial_\alpha u_{m}\rangle
\;,
\label{eq:Owg}
\end{equation}
where every object in each of these equations should consistently
carry either a (W) or (H) label.
(We have now suppressed the $\k$ subscripts
and introduced the notation $\partial_\alpha=\partial/\partial k_\alpha$ for
conciseness.)  In this notation, Eq.~(\ref{eq:omsum}) becomes
\begin{equation}
  \Omega_{\alpha\beta}(\k)=\sum_{n=1}^M f_n(\k)\,
  \Omega_{nn,\alpha\beta}\ph(\k)
\;.
\label{eq:omtot}
\end{equation}
This matrix is antisymmetric in the Cartesian indices.
Note that when $\Omega_{\alpha\beta}$ appears without a
(W) or (H) superscript, as on the left-hand side of this equation,
it denotes the total Berry curvature on the left-hand side
of Eq.~(\ref{eq:omsum}).

The matrix representation of an ordinary operator such as the
Hamiltonian or the velocity can be transformed from the Wannier to the 
Hamiltonian
gauge, or vice versa, just by operating on the left and right by
$U^\dagger(\k)$ and $U(\k)$, as in Eq.~(\ref{eq:Htrans});
such a matrix is called ``gauge-covariant.''
Unfortunately, the matrix objects in Eqs.~(\ref{eq:Awg}-\ref{eq:Owg})
are not gauge-covariant, because they involve $k$-derivatives acting on
the Bloch states. For example, a straightforward calculation shows that
\begin{equation}
A_{\alpha}\ph=
    U^\dagger A_{\alpha}\pw U
  + i U^\dagger \,\partial_\alpha U
\label{eq:Atrans}
\end{equation}
where each object is an $M\times M$ matrix and matrix products are
implied throughout.  For every matrix object ${\cal O}$, we define
\begin{equation}
\overline{\cal O}^{(\rm H)}=
U^\dagger {\cal O}^{(\rm W)}U
\label{eq:utrans}
\end{equation}
so that, by definition,
$\overline{\cal O}^{(\rm H)}={\cal O}^{(\rm H)}$
only for gauge-covariant objects.

The derivative $\partial_\alpha U$ may be obtained from ordinary
perturbation theory.  We adopt a notation in which
$\vt\phi_m\ra$ is the $m$-th $M$-component column vector of
matrix $U$, so that 
$\la\phi_n\vt H\pw \vt\phi_m\ra={\cal E}_n\,\delta_{nm}$;
the stylized bra-ket notation is used to emphasize that objects like
$H\pw$ and $\vt\phi_n\ra$ are $M\times M$ matrices and $M$-component
vectors, i.e., operators and state vectors in the ``tight-binding
space'' defined by the WFs, not in the original Hilbert space.
Perturbation theory with respect to the parameter $\k$ takes the form
\begin{equation}
\vt\partial_\alpha\phi_n\ra=\sum_{l\not= n}
  \frac{\la\phi_l\vt H_\alpha\pw\vt\phi_n\ra}{{\cal E}_n\ph-{\cal E}_l\ph}
  \,\vt\phi_l\ra
\label{eq:pert}
\end{equation}
where $H_\alpha\pw\equiv\partial_\alpha H\pw$.  In matrix notation
this can be written
\begin{equation}
\partial_\alpha U_{mn}=\sum_l\,U_{ml}\,\D_{ln,\alpha} =  (U\D_\alpha)_{mn}
\label{eq:dau}
\end{equation}
where
\begin{equation}
\D_{nm,\alpha}\equiv (U^{\dagger}
\partial_{\alpha}U)_{nm}=
\begin{cases}
  \displaystyle
  \frac{\overline H_{nm,\alpha}^{(\rm H)}}{{\cal E}^{(\rm H)}_{m}
  -{\cal E}_{n}^{(\rm H)}}& \text{if $n\not= m$}\\ \\
  0& \text{if $n=m$}
\end{cases}
\label{eq:ddef}
\end{equation}
and $\overline H_{nm,\alpha}^{(\rm H)}=( U^{\dagger}
H_{\alpha}^{(\rm W)}U)_{nm}$ according to Eq.~(\ref{eq:utrans}).
Note that while $\Omega_{\alpha\beta}$ and $A_\alpha$ are Hermitian
in the band indices, $\D_\alpha$ is instead antihermitian.
The gauge choice implicit in Eqs.~(\ref{eq:pert}) and (\ref{eq:ddef}) is
$ \la \phi_n\vt\partial_\alpha\phi_n\ra=
(U^{\dagger}\partial_{\alpha}U)_{nn}=0$ (this is the so-called ``parallel
transport'' gauge).

Using Eq.~(\ref{eq:dau}), Eq.~(\ref{eq:Atrans}) becomes
\begin{equation}
A_\alpha\ph=\overline{A}_\alpha\ph+i\D_\alpha
\label{eq:Ata}
\end{equation}
and the derivative of Eq.~(\ref{eq:twist}) becomes
\begin{equation}
|\partial_{\alpha}u_n^{(\rm H)}\rangle=
\sum_{m}|\partial_{\alpha} u_m^{(\rm W)}\rangle U_{mn}
+\sum_{m}|u_m^{(\rm H)}\rangle
\D_{mn,\alpha} \;.
\label{eq:udtrans}
\end{equation}
Plugging the latter into Eq.~(\ref{eq:Owg}), we finally obtain, after a
few manipulations, the matrix equations
\begin{eqnarray}
\Omega_{\alpha\beta}\ph &=&
\overline\Omega_{\alpha\beta}\ph - [\D_\alpha,\overline A_\beta\ph]
\nonumber\\
&&\quad + [\D_\beta,\overline A_\alpha\ph] -i[\D_\alpha,\D_\beta]\;.
\label{eq:om-a}
\end{eqnarray}
The band-diagonal elements $\Omega_{nn,\alpha\beta}\ph(\k)$ then
need to be inserted into Eq.~(\ref{eq:omtot}).

\subsection{Discussion}
\label{sec:disc}

We expect, based on Eq.~(\ref{eq:kubo}), that the largest
contributions to the AHC will come from regions of $k$-space where there
are small energy splittings between bands
(for example, near spin-orbit-split avoided crossings).\cite{fang03}
In the present formulation, this will
give rise to small energy denominators in Eq.~(\ref{eq:ddef}),
leading to very large $\D_\alpha$ values in those regions.
These large and spiky contributions will then propagate into
$A_\alpha\ph$ and $\Omega_{\alpha\beta}\ph$,
whereas $A_\alpha\pw$ and $\Omega_{\alpha\beta}\pw$,
and also $\overline A_\alpha\ph$ and $\overline\Omega_{\alpha\beta}\ph$,
will remain with their typically smaller values.
Thus, these spiky contributions will be present in the second and third
terms, and especially in the fourth term, of Eq.~(\ref{eq:om-a}).
The contributions of these various terms are illustrated for the
case of bcc Fe in Sec.~\ref{sec:berrycurv}, and we show there that
the last term typically makes by far the dominant contribution,
followed by the second and third terms, and then by the first
term.

The dominant fourth term can be recast in the form of a Kubo
formula as
\begin{equation}
-2{\rm Im}\sum_{m\ne n}
\frac{ \la \phi_{n\k} \vt H_{\alpha}^{(\rm W)} \vt \phi_{m\k} \ra
\la \phi_{m\k} \vt H_{\beta}^{(\rm W)} \vt \phi_{n\k} \ra }
{\big({\cal E}_{m}\ph-{\cal E}_{n}\ph\big)^2}\;.
\label{eq:kubom}
\end{equation}
The following differences between this equation and the 
true Kubo formula, Eq.~(\ref{eq:kubo}), should however be kept in mind.
First, the summation in 
Eq.~(\ref{eq:kubom}) is
restricted to the $M$-band projected band structure. Second, above 
$E_{\rm win}$ the projected bandstructure deviates from the original
{\it ab-initio} one. Third, even below $E_{\rm win}$, where they do
match exactly, the ``tight-binding velocity matrix 
elements'' appearing in Eq.~(\ref{eq:kubom}) differ from the 
{\it ab-initio} ones, given by Eq.~(\ref{eq:vel}). (The relation between them
is particularly simple within the inner window, and follows from 
combining the identity
$A_{nm,\alpha}=i\langle\psi_n|\hat v_\alpha|
\psi_m\rangle/(\omega_m-\omega_n)$, valid for $m\not= n$,
with Eqs.~(\ref{eq:ddef}-\ref{eq:Ata}).)
All these differences are however exactly compensated by
the previous three terms in Eq.~(\ref{eq:om-a}).
We emphasize that all terms in that equation are
defined strictly within the projected space spanned by the Wannier
functions.

We note in passing that it is possible to rewrite Eq.~(\ref{eq:om-a})
in such a way that the large spiky contributions are isolated into
a single term.  This alternative formulation, which turns out to be
related to a gauge-covariant 
curvature tensor, will be described in Appendix A.

\subsection{Sum over occupied bands}
\label{sec:bandsum}

In the above, we have proposed to compute $\Omega\ph_{nn,\alpha\beta}$
from Eq.~(\ref{eq:om-a}) and insert it into the band sum, Eq.~(\ref{eq:omtot}),
in order to compute the AHC.  However, this approach has a shortcoming
in that small splittings (avoided crossings) between a pair of
{\it occupied} bands $n$ and $m$ leads to large values of $\D_{nm,\alpha}$,
and thus to large but canceling contributions to the AHC coming from
$\Omega\ph_{nn,\alpha\beta}$ and  $\Omega\ph_{mm,\alpha\beta}$.  Here,
we rewrite the total Berry curvature (\ref{eq:omtot})
in  such a way that the cancellation is explicit.

Inserting Eq.~(\ref{eq:om-a}) into Eq.~(\ref{eq:omtot}) and interchanging
dummy labels $n\leftrightarrow m$ in certain terms, we obtain
\begin{eqnarray}
\Omega_{\alpha\beta}(\k)&=&\sum_n f_n\,\overline{\Omega}_{nn,\alpha\beta}\ph
\nonumber\\
&+& \sum_{nm} (f_m-f_n)\left(\D_{nm,\alpha}\overline{A}_{mn,\beta }\ph\right.
\nonumber\\
&&\;\;\left.               -\D_{nm,\beta }\overline{A}_{mn,\alpha}\ph
                      +i\D_{nm,\alpha}\D_{mn,\beta}\right) .
\label{eq:bsum}
\end{eqnarray}
The factors of $(f_m-f_n)$ insure that terms arising from pairs
of fully occupied states give no contribution.  Thus,
the result of this reformulation is that individual terms in
Eq.~(\ref{eq:bsum}) have
large spiky contributions only when avoided crossings or
near-degeneracies occur across the Fermi energy.  This approach is
therefore preferable from the point of view of numerical stability,
and it is the formula that we have implemented in the current work.

As expected from the discussion in Sec.~\ref{sec:disc}
and shown later in Sec.~\ref{sec:intanom}, the dominant term
in Eq.~(\ref{eq:bsum}) is the last one,
\begin{equation}
\Omega^{DD}_{\alpha\beta}=
    i \sum_{nm} (f_m-f_n) \D_{nm,\alpha}\D_{mn,\beta}
\label{eq:OmegaDD}
\end{equation}
or, in a more explicitly Kubo-like form,
\begin{equation}
  \Omega^{DD}_{\alpha\beta} = i \sum_{nm} (f_m-f_n) \; \frac
     { \overline{H}\ph_{nm,\alpha} \overline{H}\ph_{mn,\beta} }
     {\big({\cal E}_{m}\ph-{\cal E}_{n}\ph\big)^2}
\;.
\label{eq:kubototm}
\end{equation}
In the zero-temperature limit, the latter can easily be cast into a
form like Eq.~(\ref{eq:kubom}), but with the a double sum running
over occupied bands $n$ and unoccupied bands $m$, very
reminiscent of the original Kubo formula in 
Eq.~(\ref{eq:kubotot}).
We remark that $(m_e/\hbar)\overline{H}\ph_{nm,\alpha}$ coincides with the
``effective tight-binding momentum operator'' defined in 
Ref.~\onlinecite{graf95}.

It is worth pointing out that Eq.~(\ref{eq:kubom}) can be
cast explicit as a Berry curvature, the tight-binding-space analog of
Eq.~(\ref{eq:bcurv}),
\begin{equation}
\Omega_{n,\alpha\beta}^{DD}=
-2\,{\rm Im}\,\la \partial_\alpha \phi_{n\k}
\vt\partial_\beta \phi_{n\k}\ra\;.
\label{eq:bcurv_tb}
\end{equation}
In this way Eq.~(\ref{eq:kubototm}) can be written in a form that closely
resembles the total Berry curvature, Eq.~(\ref{eq:omsum}):
\begin{equation}
\Omega^{DD}_{\alpha\beta}=\sum_{n=1}^{M} f_n \,
  \Omega_{n,\alpha\beta}^{DD}\;.
\label{eq:omsum_tb}
\end{equation}

\subsection{Evaluation of the Wannier-gauge matrices}
\label{sec:wangauge}

Eq.~(\ref{eq:bsum}) is our primary result.  To review,
recall that this is a condensed notation expressing the $M\times M$
matrix $\Omega_{nm,\alpha\beta}\ph(\k)$ in terms of matrices the
$\overline{\Omega}_{nm,\alpha\beta}\ph(\k)$, etc.
The basic ingredients needed are the four matrices
$H\pw$, $H_\alpha\pw$, $A_\alpha\pw$, and $\Omega_{\alpha\beta}\pw$
at a given $\k$. Diagonalization of the first of them yields the
energy eigenvalues needed to find the occupation factors $f_n$. It also 
provides the gauge transformation $U$ which is then used
to construct $\overline{H}_\alpha\ph$,
$\overline{A}_\alpha\ph$, and $\overline{\Omega}_{\alpha\beta}\ph$
from the other three objects 
via Eq.~(\ref{eq:utrans}).  Finally, $\overline{H}_\alpha\ph$ is
inserted into Eq.~(\ref{eq:ddef}) to obtain $\D_\alpha$, and all
terms in Eq.~(\ref{eq:bsum}) are evaluated.

In this section we explain how to obtain
the matrices $H^{(\rm W)}(\k)$, $H_\alpha^{(\rm W)}(\k)$,
$A_\alpha^{(\rm W)}(\k)$ and $\Omega_{\alpha\beta}^{(\rm W)}(\k)$
at an arbitrary point $\k$ for use in the subsequent calculations
described above.

\subsubsection{Fourier transform expressions}
\label{sec:fourier}

The four needed quantities can be expressed as follows:
\begin{equation}
H_{nm}\pw(\k)=\sum_\R e^{i\k\cdot\R}\;\langle{\bf 0}n|\hat{H}|\R m\rangle\,,
\label{eq:uu}
\end{equation}
\begin{equation}
H_{nm,\alpha}\pw(\k)=\sum_\R e^{i\k\cdot\R}\;
   iR_\alpha\,\langle{\bf 0}n|\hat{H}|\R m\rangle\,,
\label{eq:vv}
\end{equation}
\begin{equation}
A_{nm,\alpha}\pw(\k)=\sum_\R e^{i\k\cdot\R}\;
   \langle{\bf 0}n|\hat{r}_\alpha|\R m\rangle\,,
\label{eq:ww}
\end{equation}
\begin{eqnarray}
\Omega_{nm,\alpha\beta}\pw(\k)&=&\sum_\R e^{i\k\cdot\R}\;\Big(
   iR_\alpha\,\langle{\bf 0}n|\hat{r}_\beta |\R m\rangle \nonumber\\
  &&\qquad -iR_\beta \,\langle{\bf 0}n|\hat{r}_\alpha|\R m\rangle \Big)
\;.
\label{eq:xx}
\end{eqnarray}
(The notation $|{\bf 0}n\rangle$ refers to the $n$'th WF
in the home unit cell $\R={\bf 0}$.)  Eq.~(\ref{eq:uu}) follows by
combining Eqs.~(\ref{eq:blochW}) and (\ref{eq:hamW}), while
Eq.~(\ref{eq:ww}) follows by combining Eqs.~(\ref{eq:blochW}) and
(\ref{eq:Awg}).  Eqs.~(\ref{eq:vv}) and (\ref{eq:xx}) are then
obtained from (\ref{eq:uu}) and (\ref{eq:ww}) using
$H_{nm,\alpha}=\partial_\alpha H_{nm}$ and
$\Omega_{nm,\alpha\beta}=\partial_\alpha A_{nm,\beta }
-\partial_\beta A_{nm,\alpha}$, respectively.

It is remarkable that the only real-space matrix elements that are
required between WFs are those of the four operators
$\hat{H}$ and $\hat{r}_\alpha$ ($\alpha=x$, $y$, and $z$).
Because the WFs are strongly localized, these matrix elements
are expected to decay rapidly as a function of lattice vector $\R$,
so that only a modest number of these real-space matrix elements
need to be computed and stored once and for all.  Collectively, they define our
``exact tight-binding model'' and suffice to allow subsequent calculation
of all needed quantities.  Furthermore, the short range
of these matrix elements in real space insures that the Wannier-gauge
quantities on the left-hand sides of Eqs.~(\ref{eq:uu}-\ref{eq:xx}) will
be smooth functions of $\k$, thus justifying the earlier discussion
in which it was argued that these objects should have no rapid variation or
enhancement in $k$-space regions where avoided crossings occur.
(Recall that such large, rapidly-varying contributions only appear
in the $\D$ matrices and in quantities that depend upon them.)
It should however be kept in mind that Eq.~(\ref{eq:bsum}) is not written
directly in terms of the smooth quantities (\ref{eq:uu}-\ref{eq:xx}), but
rather in terms of those quantities transformed
according to Eq.~(\ref{eq:utrans}). The resulting objects
are not smooth, since the matrices $U$ change rapidly with $\k$. However,
even while not smooth, they remain small.

\subsubsection{Evaluation of real-space matrix elements}
\label{sec:evalreal}

We conclude this section by discussing the calculation of
the fundamental matrix elements $\langle{\bf 0}n|\hat{H}|\R m\rangle$
and $\langle{\bf 0}n|\hat{r}_\alpha|\R m\rangle$.  There are several
ways in which these could be computed, and the choice could well vary
from one implementation to another.  One possibility would be to
construct the WFs in real space, say on a real-space grid, and then
to compute the Hamiltonian and position-operator matrix elements
directly on that grid.  In the context of a code that
uses a real-space basis (e.g., localized orbitals or grids), this
might be the best choice.  However, in the context of plane-wave
methods it is usually more convenient to work in reciprocal space
if possible.  This is in the spirit of the
Wannier-function construction scheme,\cite{marzari97,souza01}
which is formulated as a post-processing step after a conventional
{\it ab-initio} calculation carried out on a uniform $k$-point grid.
(In the following we will use the symbol $\q$ to denote the points
of this {\it ab-initio} mesh, to distinguish them from arbitrary
or interpolation-grid points denoted by $\k$.)

The end result of the Wannier-construction step are $M$ Bloch-like
functions $|u_{n\q}\pw\rangle$ at each $\q$. The WFs are obtained from
them via a discrete Fourier transform:
\begin{equation}
\label{eq:wf}
|\R n\rangle=\frac{1}{N_q^3}
\sum_\q\, e^{-i\q\cdot(\R-\hat\r)}|u_{n\q}\pw\rangle \;.
\end{equation}
This expression follows from inverting Eq.~(\ref{eq:blochW}). If the
{\it ab initio} mesh contains
$N_q\times N_q\times N_q$ points, the resulting WFs are
really periodic functions over a supercell of dimensions
$L\times L\times L$, where $L=N_q a$ and $a$ is the lattice constant of 
the unit cell. The idea then is to choose $L$ large enough that the rapid
decay of the localized WFs occurs on a scale much smaller than $L$. 
This ensures that the matrix
elements
$\langle{\bf 0}n|\hat{H}|\R m\rangle$ and
$\langle{\bf 0}n|\hat{r}_\alpha|\R m\rangle$
between a pair of WFs separated by more than $L/2$ are negligible,
so that further refinement of the {\it ab-initio} mesh will have
a negligible impact on the accuracy of Wannier-interpolated
quantities. (In particular, the interpolated band structure,
Fig.~\ref{fig:band}, is able to reproduce tiny features of
the full bandstructure, such as spin-orbit-induced avoided
crossings, even if they occur on a length scale much smaller
than the {\it ab-initio} mesh spacing.)  While the choice of
reciprocal-space cell spanned by the vectors $\q$ is immaterial,
because of the periodicity of reciprocal space, this is not so
for the vectors $\R$. In practice we choose the $N_q\times
N_q\times N_q$ vectors $\R$ to be evenly distributed
on the Wigner-Seitz supercell of volume $N_q^3 a^3$ centered
around $\R=\bf 0$.\cite{souza01} This is the most isotropic choice
possible, ensuring that the strong decay of the matrix elements for 
$|\R|\sim L/2$ is achieved irrespective of direction.

The matrix elements of the Hamiltonian are obtained from Eq.~(\ref{eq:wf})
as
\begin{eqnarray}
\langle{\bf 0}n|\hat{H}|\R m\rangle= \frac{1}{N_q^3}\sum_\q\,
              e^{-i\q \cdot \R}H\pw_{nm}(\q)\;,
\label{eq:hw}
\end{eqnarray}
which is the reciprocal of Eq.~(\ref{eq:uu}), with the sum running
over the coarse {\it ab-initio} mesh.  The position
matrix is obtained similarly by inverting  Eq.~(\ref{eq:ww}):
\begin{equation}
\langle {\bf 0} n|\hat r_{\alpha}|\R m\rangle =\frac{1}{N_q^3}\sum_\q\,
e^{-i\q \cdot \R} A_{nm,\alpha}\pw(\q) \;.
\end{equation}
The matrix $A_{nm,\alpha}\pw(\q)$ is then evaluated by
approximating the $k$-derivatives in Eq.~(\ref{eq:Awg}) by finite-differences 
on the {\it ab-initio} mesh using the expression\cite{marzari97}
\begin{equation}
        A_{nm,\alpha}\pw(\q)\simeq
        i\sum_\b\,
        w_b b_{\alpha}\Big ( \langle u_{n\q}\pw|u_{m,\q+\b}\pw
        \rangle
        - \delta_{nm}\Big )\;,
\label{eq:A_dis}
\end{equation}
where $\b$ are the vectors connecting $\q$ to its nearest
neighbors on the {\it ab-initio} mesh. This approximation is valid because
the Bloch states vary smoothly with $\k$ in the Wannier gauge.
We note that the overlap matrices appearing on the right-hand side are
available ``for free'' as they have already been computed and stored
during the WF construction procedure.
This is also the case for the matrices 
$H\pw(\q)$ needed in Eq.~(\ref{eq:hw}).

It should be noted that the $k$-space finite-difference procedure outlined
above entails an error of order ${\cal O}(\Delta q^2)$ in the
values of the position operator matrix elements, where $\Delta q$
is the {\it ab-initio} mesh spacing.  The importance of such an error
is easily assessed by trying denser $q$-point meshes; in our case, we find
that it is not a numerically significant source of error for the
$8\times8\times8$ mesh that we employ in our calculations.
(In large measure this is simply because less than
0.5\% of the total AHC comes from terms that depend on these
position-operator matrix elements, as will be discussed in
Section~\ref{sec:results}.  Indeed, we find that the
${\cal O}(\Delta q^2)$ convergence of this small contribution hardly
shows in the convergence of the total AHC, which
empirically appears to be approximately exponential in the {\it
ab-initio} mesh density.)
However, if the ${\cal O}(\Delta q^2)$ convergence is a source of
concern, one could adopt the direct
real-space mesh integration method mentioned at the beginning of this
subsection, which should be free of such errors.

\section{Computational details}
\label{sec:cd}

In this section we present some of the detailed steps of the
calculations as they apply to our test system of bcc Fe.  First, we
describe the first-principles bandstructure calculations that are
carried out initially.
Second, we discuss the procedure for
constructing maximally localized Wannier functions for the bands of
interest following the method of Souza, Marzari, and Vanderbilt.
\cite{souza01} Third, we discuss the variable treatment of
the spin-orbit interaction within these first-principles
calculations, which is useful for testing the dependence of the AHC on the
spin-orbit coupling.

\subsection{Band structure calculation}

Fully relativistic band structure calculations for bcc Fe in
its ferromagnetic ground state at the experimental lattice constant
$a=5.42$\,Bohr are carried out using the {\tt PWSCF}
code.\cite{pwscf} A kinetic-energy cutoff of 60 Hartree is used
for the planewave expansion of the valence wavefunctions (400
Hartree for the charge densities).  Exchange and correlation
effects are treated with the PBE generalized-gradient
approximation.\cite{perdew96}

The core-valence interaction is described here by means of
norm-conserving pseudopotentials which include spin-orbit
effects\cite{soc2,dalcorso05} 
in separable Kleinman-Bylander form.
(Our overall Wannier interpolation approach
is quite independent of this specific choice and can
easily be generalized to other kinds of pseudopotentials or to
all-electron methods.)
The pseudopotential was constructed using a reference valence
configuration of $3d^74s^{0.75}4p^{0.25}$. We treat the overlap
of the valence states with the semicore $3p$ states using the
non-linear core correction approach.\cite{nlcc} The pseudopotential
core radii for the $3d$, $4s$ and $4p$ states are $1.3$, $2.0$
and $2.2$\,Bohr, respectively.  We find the small cut-off radius
for the $3d$ channel to be necessary in order to reproduce the
all-electron bandstructure accurately.

We obtain the self-consistent ground state using a 16$\times$16$\times$16
Monkhorst-Pack\cite{mp} mesh of $k$-points and a fictitious Fermi
smearing \cite{coldsmear} of 0.02\,Ry for the Brillouin-zone integration.
The magnetization is along the [001] direction, so that
the only non-zero component of the total Berry curvature is the one
along $z$.  The spin magnetic moment is found to be
2.22\,$\mu_{\rm B}$,
the same as that from an 
all-electron calculation\cite{yao04} and close to the experimental
value of 2.12\,$\mu_{\rm B}$. 

In order to calculate the Wannier functions, we freeze the self-consistent
potential and perform a non-self-consistent calculation on a uniform
$n\times n\times n$ grid of $k$-points.  We tested several grid densities
ranging from $n$=4 to $n$=10 and ultimately chose $n$=8 (see end of
next subsection).  Since we want to construct 18 WFs ($s$, $p$, and
$d$-like for spin up and down), we need to include a sufficient number of
extra bands to cover the orbital character of these intended
WFs everywhere in the Brillouin zone.  With this in mind, we
calculate the first 28 bands at each $k$-point, and then exclude
any bands above 58\,eV (the ``outer window'' of
Ref.~\onlinecite{souza01}).  The 18 WFs are then disentangled from
the remaining bands using the procedure described in the next
section.

\subsection{Maximally-localized spinor Wannier functions for bcc Fe}
\label{sec:maxloc}

The energy bands of interest (extending up to, and just above,
the Fermi energy) have mainly mixed $s$ and $d$ character and
are entangled with the bands at higher energies.  In order to
construct maximally-localized WFs to describe these bands,
we use a two-step post-processing procedure\cite{souza01}
as implemented in the {\tt WANNIER90} code.\cite{wannier} 
In the first (``disentangling'') step,
an 18-band subspace (the ``projected space'') is identified that
minimizes the invariant part of the spread functional, subject
to the constraint of including the states within an  
inner energy window.\cite{souza01} We chose this window to span an energy
range of 30~eV from the bottom of the valence bands
(up to $E_{\rm win}$ in Fig.~\ref{fig:band}).
In the second step,
a set of maximally-localized WFs spanning this
subspace is chosen by minimizing the gauge-dependent part of the 
spread functional.\cite{marzari97}

Although the original prescription for obtaining maximally-localized Wannier
functions was formulated for the spinless case, it is trivial to
adapt it to treat spinor wavefunctions, in which case the resulting WFs also
have spinor character: each element of the overlap matrix,
which is the key input to the WF-generation code, is
simply calculated as the sum of two spin components,
\begin{equation}
  S_{\k,\b}^{nm} = \sum_{\sigma} \,
  \langle u_{n\k}^{\sigma} | u_{m,\k+\b}^{\sigma} \rangle \;,
\end{equation}
where $| u_{n,\k+\b}^{\sigma} \rangle$ is one of the two components
of the spinor wavefunction. 

In order to facilitate later analysis (e.g., of the orbital and
spin character of various bands), we have modified the second
step as follows.  At each $k$-point on the $8\times 8\times 8$
mesh, we form the 18$\times$18 matrix representation of the spin
operator $\hat S_z=(\hbar/2)\hat\sigma_z$ 
in the space of band states and diagonalize it.
The $18$-dimensional space at this $k$-point is then divided in
two $9$-dimensional subspaces, a mostly spin-up subspace spanned
by the eigenstates having $S_z$ eigenvalues close to $+1$, and
a mostly spin-down subspace associated with eigenvalues close
to $-1$ (we will use units of
$\hbar/2$ whenever we discuss $S_z$ in the remainder of the manuscript).  
The spread functional is then minimized within each of
these subspaces separately.  We thus emerge with 18 well-localized
WFs divided into two groups: nine that are almost entirely spin-up
and nine that are almost entirely spin-down (in practice we find
$|\langle \hat S_z \rangle|> 0.999$ in all cases).
While this procedure results in a total spread that is slightly
greater than would be obtained otherwise, we find that the difference
is very small in practice,
and the imposition of these rules makes for a much more transparent
analysis of subsequent results.  For example, it makes it much easier
to track the changes in the WFs before and after the spin-orbit
coupling is turned on, or to identify the spin character
of various pieces of the Fermi surface.

\begin{figure}
\begin{center}
\epsfig{file=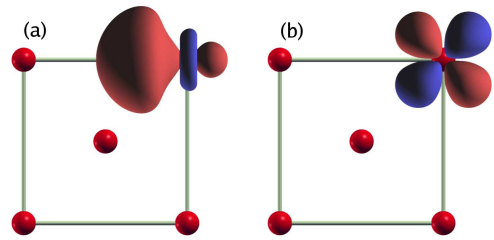,width=2.6in}
\end{center}
\caption{(Color online). 
Isosurface contours of maximally-localized spin-up WF in
bcc Fe (red for positive value and blue for negative value), for
the $8 \times 8 \times 8$ $k$-point sampling. (a) $sp^3d^2$-like WF
centered on a Cartesian axis; (b) $d_{xy}$-like WF centered on the
atom.}
\label{fig:wfs}
\end{figure}

To start the minimization procedure, we choose trial functions having
pure spin character (up or down) and a spatial form of
a Gaussian times a predetermined angular factor.
In our first attempts, we chose angular factors appropriate for the
three $t_{2g}$ states $d_{xy}$, $d_{xz}$, and 
$d_{yz}$; the two $e_g$ states
$d_{z^2}$ and $d_{x^2-y^2}$; the three $p$ states $p_x$, $p_y$, and
$p_z$; and $s$.  The iterative procedure\cite{souza01}
then projects these onto
the band subspace and improves upon them.  We found that the spread
minimization procedure converted the $t_{2g}$ trial functions into
$t_{2g}$-like WFs, while it mixed the other six states to form six
hybrid WFs of $sp^3d^2$-type.\cite{Pauling31}
Having discovered this, we have modified our procedure accordingly:
henceforth, we choose three $t_{2g}$-like trial functions and
six $sp^3d^2$-like ones.  With this initialization, we find the
convergence to be quite rapid, with only about 100 iterations
needed to get a well-converged spread functional.

The WFs that result from this procedure are shown in Fig.~\ref{fig:wfs}.
The up-spin WFs are plotted,
but the WFs are very similar for both spins.
An example of an $sp^3d^2$-hybrid WF is shown in
Fig.~\ref{fig:wfs}(a); this one extends along the $-x$ axis,
and the five others are similarly projected along the $+x$, $\pm y$,
and $\pm z$ axes.  One of the $t_{2g}$-like WFs is shown in
Fig.~\ref{fig:wfs}(b); this one has $xy$ symmetry, while the others
have $xz$ and $yz$ symmetry.  The centers of the $sp^3d^2$-like WFs
are slightly shifted from the atomic center along $\pm x$, $\pm y$,
or $\pm z$, while the  $t_{2g}$-like WFs remain centered on the atom.

We studied the convergence of the WFs and interpolated bands as a function
of the density $n\times n\times n$ of the Monkhorst-Pack $k$-mesh
used for the initial {\it ab-initio} calculation.  We tested
$n=4$, 6, 8, and 10, and found that $n=8$ provided the best tradeoff
between interpolation accuracy and computational cost.  This is
the mesh that was used in generating the results presented in
Sec.~\ref{sec:results}.

\subsection{Variable spin-orbit coupling in the pseudopotential framework}
\label{sec:spinorbit}

Since the AHE present in ferromagnetic iron is a spin-orbit-induced effect,
it is obviously important
to understand the role of this coupling as thoroughly as possible.
For this purpose, it is very convenient to be able to treat the
strength of the coupling as an adjustable parameter.  For example,
by turning up the spin-orbit coupling continuously from
zero and tracking how various contributions to the AHC behave, it
is possible to separate out those contributions that are of linear,
quadratic, or higher order in the coupling strength. 
Some results of this kind will be given later in Sec.~\ref{sec:results}.

Because the spin-orbit coupling is a relativistic effect, it is
appreciable mainly in the core region of the atom where the
electrons have relativistic velocities.  In a pseudopotential framework
of the kind adopted here, both the scalar relativistic effects and
the spin-orbit coupling are included in the pseudopotential construction.
For example, in the Bachelet-Hamann semilocal pseudopotential scheme,
\cite{bachelet82} the construction procedure generates, for each
orbital angular momentum $l$, a scalar-relativistic potential
$V^{\rm sr}_l(r)$ and a spin-orbit difference potential
$V^{\rm so}_l(r)$ which enter the Hamiltonian in the form
\begin{equation}
\hat{V}_{\rm ps}=\sum_l \hat{P}_l\,\left[ V^{\rm sr}_l(r) + \lambda\,
    V^{\rm so}_l(r)\, {\bf L}\cdot {\bf S}\right] \;,
\end{equation}
where $\hat{P}_l$ is the projector onto states of orbital angular
momentum $l$ and $\lambda$ controls the strength of spin-orbit coupling
(with $\lambda$=1 being the physical value).  For the free atom,
this correctly leads to eigenstates labeled by total angular
momentum $j=l\pm1/2$.

In our calculations, we employ fully non-local pseudopotentials instead
of semilocal ones because of their computationally efficient form.
In this case, controlling the strength of the spin-orbit coupling
requires some algebraic manipulation.
We write the norm-conserving non-local pseudopotential operator as
\begin{equation}
\hat{V}_{\rm ps}=|\beta_{lj\mu}\rangle\, D_{lj} \,\langle\beta_{lj\mu}|
\label{eq:Vps}
\end{equation}
where there is an implied sum running over the indices (orbital angular
momentum $l$, total angular momentum $j=l\pm1/2$, and $\mu=-j,...,j$)
and species and atomic position indices have been suppressed.
The $|\beta_{lj\mu}\rangle$ are radial
functions multiplied by appropriate spin-angular functions and the $D_{lj}$
are the channel weights.
We introduce the notation
$\beta^{(+)}_{l}(r)$ and $\beta^{(-)}_{l}(r)$ for the radial
parts of $|\beta_{l,l+1/2,\mu}\rangle$ and $\beta_{l,l-1/2,\mu}\rangle$,
respectively, and similarly define $D^{(\pm)}_{l}=D_{l,l\pm1/2}$.
Using this notation, we can define the scalar-relativistic
(i.e., $j$-averaged) quantities
\begin{equation}
        D^{\rm sr}_{l}=\frac{l+1}{2l+1}\,D^{(+)}_l
                         +\frac{l}{2l+1}\,D^{(-)}_l  \;,
\end{equation}
\begin{equation}
 \beta^{\rm sr}_{l}(r)=
   \frac{l+1}{2l+1}\,
      \sqrt{\frac{D^{(+)}_l}{D^{\rm sr}_{l}}}\;\,\beta^{(+)}_l(r)
  +\frac{l}{2l+1}\,
      \sqrt{\frac{D^{(-)}_l}{D^{\rm sr}_{l}}}\; \,\beta^{(-)}_l(r)
\end{equation}
and the corresponding spin-orbit difference quantities
\begin{equation}
D^{\rm so}_{lj}=D_{lj}-D^{\rm sr}_{l}\;,
\end{equation}
 \begin{equation}
 |\beta^{\rm so}_{lj\mu}\rangle=
    |\beta_{lj\mu}\rangle-|\beta^{\rm sr}_{lj\mu}\rangle \;.
 \end{equation}
 where $|\beta^{\rm sr}_{lj\mu}\rangle$ is $\beta^{\rm sr}_{l}(r)$
 multiplied by the spin-angular function with labels $(lj\mu)$.
Then the non-local pseudopotential can be written as
\begin{equation}
\hat{V}_{\rm ps}=\hat{V}^{\rm sr}+\lambda\,\hat{V}^{\rm so}
\end{equation}
where
\begin{equation}
\hat{V}^{\rm sr}=|\beta^{\rm sr}_{lj\mu}\rangle \, D^{\rm sr}_{l} \,
    \langle \beta^{\rm sr}_{lj\mu} |
\label{eq:Vsr}
\end{equation}
and
\begin{eqnarray}
\hat{V}_{\rm so}&=&|\beta^{\rm sr}_{lj\mu}\rangle \, D^{\rm so}_{lj}\, 
    \langle \beta^{\rm sr}_{lj\mu} | \nonumber\\&&
     +\,|\beta^{\rm so}_{lj\mu}\rangle \, (D^{\rm sr}_{l}+D^{\rm so}_{lj}) \,
    \langle \beta^{\rm sr}_{lj\mu} | \nonumber\\&&
     +\,|\beta^{\rm sr}_{lj\mu}\rangle \, (D^{\rm sr}_{l}+D^{\rm so}_{lj}) \,
    \langle \beta^{\rm so}_{lj\mu} | \nonumber\\&&
     +\,|\beta^{\rm so}_{lj\mu}\rangle \, (D^{\rm sr}_{l}+D^{\rm so}_{lj}) \,
    \langle \beta^{\rm so}_{lj\mu} | \;.
\end{eqnarray}
This clearly reduces to the desired results (\ref{eq:Vps}) for $\lambda=1$ and
(\ref{eq:Vsr}) for $\lambda=0$.

\section{Results}
\label{sec:results}

In this section, we present the results of the calculations of the
Berry curvature and its integration over the BZ using the formulas
presented in Sec.~\ref{sec:fdbc}, for the case of bcc Fe.

\subsection{Berry Curvature}
\label{sec:berrycurv}

\begin{figure}
\begin{center}
 \epsfig{file=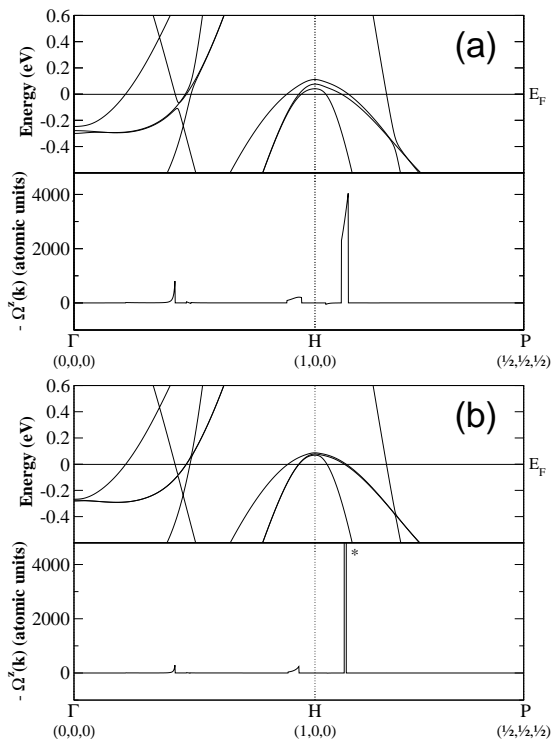,width=2.8in}
\end{center}
\caption{Band structure and total Berry curvature, as calculated
using Wannier interpolation, plotted along the path $\Gamma$--H--P
in the Brillouin zone.
(a) Computed at the full spin-orbit coupling strength $\lambda=1$.
(b) Computed at the reduced strength $\lambda=0.25$.
The peak marked with a star has a height of 5$\times$10$^4$\,a.u. }
\label{fig:bdbc}
\end{figure}

We begin by illustrating the very sharp and strong variations that
can occur in the total Berry curvature, Eq.~(\ref{eq:omega_tot}),
near Fermi-surface features in the bandstructure.\cite{fang03}  In
Fig.~\ref{fig:bdbc}(a) we plot the energy bands (top subpanel) and the total
Berry curvature (bottom subpanel) in the vicinity of the the
zone-boundary point ${\rm H}=\frac{2\pi}{a}(1,0,0)$, where three
states, split by the spin-orbit interaction, lie just above the Fermi
level. The large spike in the Berry curvature between the H and P
points arises where two bands, split by the spin orbit interaction, lie
on either side of the Fermi level.\cite{yao04} 
This gives rise to small
energy denominators, and hence large contributions, mainly in
Eq.~(\ref{eq:kubototm}).  On reducing the strength of the
spin-orbit interaction as in Fig.~\ref{fig:bdbc}(b), the
energy separation between these bands is reduced, resulting in a
significantly sharper and higher spike in the Berry curvature.
A second type of sharp structure is visible in Fig.~\ref{fig:bdbc},
where one can see two smaller spikes, one at about 40\% and another
at about 90\% of the way from $\Gamma$ to H, which
decrease in magnitude as the as the spin-orbit
coupling strength is reduced.  These arise from pairs of bands that
straddle the Fermi energy even in the absence of spin-orbit
interaction.  Thus, the small spin-orbit coupling does not shift the
energies of these bands significantly, but it does induce an
appreciable Berry curvature that is roughly linear in the spin-orbit
coupling.

\begin{figure}
\begin{center}
\epsfig{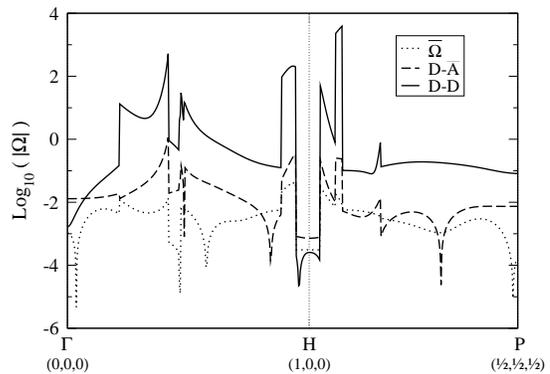} 
\end{center}
\caption{Decomposition of the total Berry curvature into contributions coming
from the three kinds of terms appearing in Eq.~(\ref{eq:bsum}).
The path in $k$-space is the same as in Fig.~\ref{fig:bdbc}.
Dotted line is the first ($\overline{\Omega}$) term, dashed line is the sum
of second and third ($D$--$\overline{A}$) terms, and solid line is the
fourth ($D$--$D$) term of
Eq.~(\ref{eq:bsum}). Note the log scale on the vertical axis.}
\label{fig:3bc}
\end{figure}

The decomposition of the total Berry curvature into its various
contributions in Eq.~(\ref{eq:bsum}) is illustrated by plotting
the first (``$\overline{\Omega}$'') term,
the second and third (``$D$--$\overline{A}$'') terms,
and the fourth (``$D$--$D$'' or Kubo-like) term of Eq.~(\ref{eq:bsum})
separately along the line $\Gamma$--H--P.  Note the logarithmic scale.
The results confirm the expectations expressed in
Secs.~\ref{sec:disc} and \ref{sec:bandsum}, namely, that the largest
terms would be those reflecting large contributions to $D$ arising
from small energy denominators.  Thus, the $\overline{\Omega}$
term remains small everywhere, the $D$--$\overline{A}$ terms become
one or two orders of magnitude larger at places where small energy
denominators occur, and the $D$--$D$ term, Eq.~(\ref{eq:kubototm}),
is another one or two orders
larger in those same regions.  Scans along other
lines in $k$-space reveal similar behavior.  We may therefore expect
that the $D$--$D$ term will make the dominant overall contribution
to the AHC. As we shall show in the next subsection, this is precisely
the case.

\begin{figure}
\begin{center}
\epsfig{file=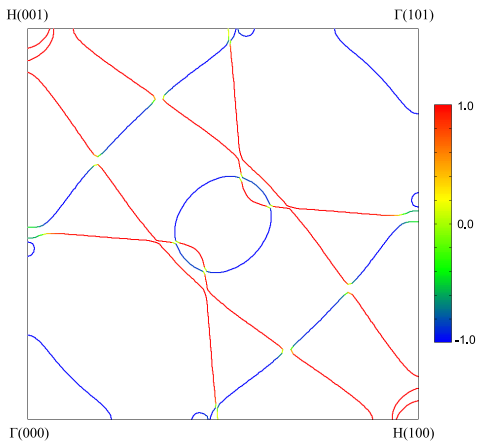,width=2.5in}
\end{center}
\caption{(Color online). 
Lines of intersection between the Fermi surface and the plane $k_y=0$.
Colors indicate the $S_z$ spin-component of the states
on the Fermi surface (in units of $\hbar/2$).}
\label{fig:fermispin}
\end{figure}

In order to get a better feel for the connection between Fermi surface
features and the Berry curvature, we next inspect these quantities
on the $k_y=0$ plane in the Brillouin zone, following Ref.~\onlinecite{yao04}.
In Fig.~\ref{fig:fermispin}
we plot the intersection of the Fermi surface with this plane
and indicate, using color coding, the $S_z$ component of the spin
carried by the corresponding wavefunctions. 
The good agreement between the shape of the Fermi surface given here
and in Fig.~3 of Ref.~\onlinecite{yao04} is further evidence that the
accuracy of our approach matches that of all-electron methods.
It is evident that the presence of the spin-orbit interaction, in
addition to the exchange splitting, is sufficient to remove all
degeneracies on this plane,\cite{singh73} 
changing significantly the connectivity of the Fermi surface.

\begin{figure}
\begin{center}
\epsfig{file=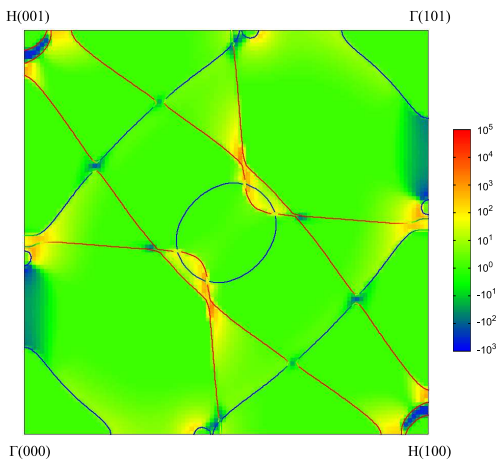,width=2.6in}
\end{center}
\caption{(Color online).
Calculated total Berry curvature $\Omega_z$ in the plane $k_y=0$
(note log scale).  Intersections of the Fermi surface with this
plane are again shown.}
\label{fig:bc}
\end{figure}

The calculated Berry curvature is shown in Fig.~\ref{fig:bc}. It
can be seen that the regions in which the Berry curvature is small
(light green regions) fill most of the plane.  The largest values
occur at the places where two Fermi lines approach one another,
consistent with the the discussion of Fig.~\ref{fig:bdbc}.
Of special importance are the avoided crossings between two
bands having the same sign of spin, or between two bands of
opposite spin.  Examples of both kinds are visible in the figure,
and both tend to give rise to very large contributions in the
region of the avoided crossing.
Essentially, the spin-orbit interaction
causes the character of these bands to change extremely rapidly with
$\k$ near the avoided crossing; this is the origin of the
large Berry curvature.  The large contributions near the H points
correspond to the peaks that were already mentioned in the discussion of
Fig.~\ref{fig:bdbc}, resulting from mixing of nearly degenerate bands
by the spin-orbit interaction.

\subsection{Integrated anomalous Hall conductivity}
\label{sec:intanom}

We now discuss the computation of the AHC as an integral
of the Berry curvature over the Brillouin zone,
Eq.~(\ref{eq:sigma_b}).  We first define a nominal $N_0\times
N_0\times N_0$ mesh that uniformly fills the Brillouin zone.
We next reduce this to a sum over the irreducible wedge
that fills $\frac{1}{16}$th of the Brillouin zone, using the
tetragonal point-group symmetry (broken from cubic by the onset
of ferromagnetism), and calculate $\Omega_z$ on each mesh point
using Eq.~(\ref{eq:bsum}).  Finally, following Yao {\it et al.},
\cite{yao04} we implement an adaptive
mesh refinement scheme in which we identify those points of
the $k$-space mesh at which the computed Berry curvature exceeds
a threshold value $\Omega_{\rm cut}$, and recompute $\Omega_z$ on
an $N_a\times N_a\times N_a$ submesh spanning the original cell
associated with this mesh point.  The AHC is then computed as a sum
of $\Omega_z$ over this adaptively refined mesh with appropriate
weights.

\begin{table}
\caption{Convergence of AHC with respect to the density of the
nominal $k$-point mesh (left column) and the adaptive refinement
scheme used to subdivide the mesh in regions of large contributions
(middle column).}
\begin{ruledtabular}
\begin{tabular}{ccc}
$k$-point mesh & Adaptive refinement & $\sigma$ $(\Omega$ ${\rm cm})^{-1}$ \\
\hline
$200 \times 200 \times 200$ & $3 \times 3 \times 3$ & 774.55 \cr
$250 \times 250 \times 250$ & $3 \times 3 \times 3$ &774.84\cr
$320 \times 320 \times 320$ & $3 \times 3 \times 3$ &775.80\cr
$200 \times 200 \times 200$ & $5 \times 5 \times 5$ &765.96\cr
$250 \times 250 \times 250$ & $5 \times 5 \times 5$ &766.37\cr
$320 \times 320 \times 320$ & $5 \times 5 \times 5$ & 766.76\cr
$200 \times 200 \times 200$ & $7 \times 7 \times 7$ &763.87\cr
$250 \times 250 \times 250$ & $7 \times 7 \times 7$ &764.84\cr
$320 \times 320 \times 320$ & $7 \times 7 \times 7$ & 765.10\cr
$320 \times 320 \times 320$ & $9 \times 9 \times 9$ & 764.59\cr
$320 \times 320 \times 320$ & $11 \times 11 \times 11$ & 764.37\cr
$320 \times 320 \times 320$ & $13 \times 13 \times 13$ &764.27\cr
\end{tabular}
\end{ruledtabular}
\label{table:convergence}
\end{table}

The convergence of the AHC with respect to the choice of
mesh is presented in Table~\ref{table:convergence}. 
We have chosen $\Omega_{\rm cut}=1.0\times10^2$~a.u., which causes the
adaptive mesh refinement to be triggered at approximately 0.11\% of
the original mesh points.
The value of 751\,$(\Omega$ ${\rm cm})^{-1}$ reported previously in
Ref.~\onlinecite{yao04} corresponds to a mesh similar to the one in
the first line of Table~\ref{table:convergence}; our value of
775\,$(\Omega$ ${\rm cm})^{-1}$ for this mesh thus agrees to within
a few percent with their value.
Based on the results of Table~\ref{table:convergence}, we estimate
the converged value of $\sigma$ to be
764\,$(\Omega$ ${\rm cm})^{-1}$.

It can be seen from Table~\ref{table:convergence}
that a $200\times200\times200$ mesh with $3\times3\times3$
refinement approaches within $\sim$1\% of the converged value.  It is
also evident that the level of refinement is more important than
the fineness of the nominal mesh; a $200\times200\times200$
mesh with $5\times5\times5$ adaptive refinement yields a result
that is within 0.1\% of the converged value, better than a
$320\times320\times320$ mesh with a lower level of refinement.

\begin{table}
\caption{Contributions to the AHC coming from different regions
of the Brillouin zone, as defined in the text.}
\begin{ruledtabular}
\begin{tabular}{cccc}
$\Delta  E$  (eV)&   like-spin (\%) & opposite-spin (\%) & smooth (\%)\\
\hline
0.1 & 21 & 26 & 53 \cr
0.2 & 23 & 51 & 26 \cr
0.5 & 30 & 68 & 2\cr
\end{tabular}
\end{ruledtabular}
\label{table:percentage}
\end{table}

It is interesting to decompose the total AHC into contributions coming
from different parts of the Brillouin zone.  For example, as we saw in
Fig.~\ref{fig:bc}, there is a smooth, low-intensity background
that fills most of the volume of the Brillouin zone, and it is hard to
know {\it a priori} whether the total AHC is dominated by these
contributions or by the much larger ones concentrated in small
regions.  With this motivation, we have somewhat arbitrarily
divided the Brillouin zone into three kinds of regions, which we
label as `smooth', `like-spin', and `opposite-spin'.  To do this,
we identify $k$-points at which there is an occupied band in the
interval $[E_f-\Delta E,E_f]$ and an unoccupied band in the interval
$[E_f,E_f+\Delta E]$, where $\Delta E$ is arbitrarily chosen to be a
small energy such as $0.1$, $0.2$, or $0.5$\,eV.  If so,
the $k$-point is said to belong to the `like-spin' or
`opposite-spin' region depending on whether the dominant characters of
the two bands below and above the Fermi energy are of the same
or of opposite spin.  Otherwise, the $k$-point is assigned to the
`smooth' region.
As shown in Table~\ref{table:percentage}, the results depend strongly
on the value of $\Delta E$. Overall, what is clear is that the major
contributions arise from the bands within $\pm 0.5$\,eV of $E_f$,
and that neither like-spin nor opposite-spin contributions are dominant.

Next, we return to the discussion of the decomposition of
the total Berry curvature in Eq.~(\ref{eq:bsum}) into
$\overline{\Omega}$, $D$--$\overline{A}$, and $D$--$D$ terms.
We find that these three kinds of terms account for $-$0.20\%,
0.71\%, and  99.48\%, respectively, of the total AHC.
(Similarly, for the alternative decomposition of
Appendix A, the second term of Eq.~(\ref{eq:bsumalt}) is
found to be responsible for more than 99\% of the total.)
Thus, if a 1\% accuracy is acceptable, one could
actually neglect the $\overline{\Omega}$ and $D$--$\overline{A}$
terms entirely, and approximate the total AHC by the $D$--$D$
(Kubo-like) terms alone, Eq.~(\ref{eq:kubototm}).  

While we had anticipated in Sec.~\ref{sec:disc}
that the $D$--$D$ terms should be expected to dominate, the extent to which 
that occured in the actual calculation is somewhat surprising and merits 
further discussion. It is important to emphasize that this should
not be expected to occur when using an arbitrary Wannier representation, but 
only for WFs which minimize the spread functional. Indeed, only the sum of all 
terms in Eq.~(\ref{eq:bsum}) is uniquely defined; taken separately, the 
$\overline{\Omega}$, $D$--$\overline{A}$, and $D$--$D$ terms
depend on the choice of gauge. Moreover, while the $\overline{\Omega}$ and
$D$--$\overline{A}$ terms involve both the Hamiltonian and position matrix
elements between WFs, the dominant $D$--$D$ term only depends on the 
Hamiltonian matrix elements. Since the minimization of the gauge-dependent
part of the spread functional corresponds precisely to minimizing the RMS
average magnitude of the position matrix element between WFs,\cite{marzari97}
it is perhaps not too surprising that we capture most of the AHC by neglecting
the terms which involve position matrix elements.

From a computational point of view,
the fact that the $D$--$D$ terms are fully specified by the
Hamiltonian matrix elements alone means that considerable
savings can be obtained by avoiding the evaluation of
the Fourier transforms in Eqs.~(\ref{eq:ww}-\ref{eq:xx})
at every interpolation point
(and avoiding the setup of the matrix elements 
$\langle {\bf 0}n|\hat r_\alpha|{\bf R}m\rangle$, which can be costly in a
real-space implementation). More importantly,
this observation, if it turns out to hold for other materials as well,
could prove to be important for future efforts to
derive approximate schemes capable of capturing the most important
contributions to the AHC.

\begin{figure}
\begin{center}
\epsfig{file=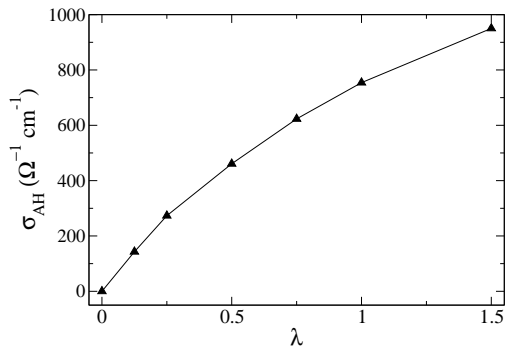,width=2.6in}
\end{center}
\caption{Anomalous Hall conductivity vs.~spin-orbit coupling strength.}
\label{fig:lam}
\end{figure}

Finally, we investigate how the total AHC depends upon the
strength of the spin-orbit interaction, following the approach
of Sec.~\ref{sec:spinorbit} to modulate the spin-orbit strength.
The result is shown in Fig.~\ref{fig:lam}.  We emphasize that our
approach is a more specific test of the dependence upon spin-orbit
strength than the one carried out in Ref.~\onlinecite{yao04};
there, the speed of light $c$ was varied, which entails changing
the strength of the various scalar relativistic terms as well.
Nevertheless, both studies lead to a similar conclusion: the
variation is found to be linear for small values of the spin-orbit
coupling ($\lambda\ll 1$), while quadratic or other higher-order
terms also become appreciable when the full interaction is included
($\lambda=1$).

\subsection{Computational Considerations}

The computational requirements for this scheme are quite modest. 
The self-consistent ground state calculation and the construction
of the WFs takes 2.5 hours on a single 2.2GHz AMD-Opteron processor.
The expense of computing the AHC as a sum over interpolation mesh
points depends strongly on the density of the mesh.  On the same processor
as above, the average CPU time to evaluate $\Omega_z$ on each
$k$-point was about 14\,msec. We find that the mesh
refinement operation does not significantly increase the total
number of $k$-point evaluations until the refinement level $N_a$
exceeds $\sim$10.  Allowing for the fact that the calculation only
needs to be done in the irreducible $\frac{1}{16}$th of the
Brillouin zone, the cost for the AHC evaluation on a
200$\times$200$\times$200 mesh is about 2 hours.

The CPU time per $k$-point evaluation is dominated (roughly 90\%)
by the Fourier transform operations needed to construct the objects
in Eqs.~(\ref{eq:uu}-\ref{eq:xx}). The diagonalization of the
18$\times$18 Hamiltonian matrix, and other operations needed to
compute Eq.~(\ref{eq:bsum}), account for only about 10\% of the
time.  The CPU requirement for the Fourier transform step is
roughly proportional to the number of $\R$ vectors kept in
Eqs.~(\ref{eq:uu}-\ref{eq:xx}); it is possible that this number
could be reduced by exploring more sophisticated methods for
truncating the contributions coming from the more
distant $\R$ vectors.

Of course, the loop over $k$-points in the AHC calculation is
trivial to parallelize, so for dense $k$-meshes we speed up this
stage of the calculation by distributing across multiple processors.

\section{Summary and Discussion}
\label{sec:conclusion}

In summary, we have developed an efficient method for computing the
intrinsic contribution to the anomalous Hall conductivity of a
metallic ferromagnet as a Brillouin-zone integral of the Berry
curvature. Our approach is based on Wannier interpolation, a powerful
technique for evaluating properties that require a very dense sampling
of the Brillouin zone or Fermi surface. The key idea is to map the low-energy
first-principles electronic structure onto an ``exact
tight-binding model'' in the basis of appropriately constructed Wannier 
functions, which are typically partially occupied.
In the Wannier representation the desired quantities can then be evaluated
at arbitrary $k$-points at very low computational cost.
All that is needed is to evaluate, once and for all, the Wannier-basis
matrix elements of the Hamiltonian and a few other
property-specific operators (namely, for the Berry curvature, the
three Cartesian position operators).

When evaluating the Berry curvature in this 
way, the summation over all unoccupied bands and the expensive
calculation of the velocity matrix elements needed in the traditional Kubo
formula are circumvented.\cite{explan-kubo}
They are replaced by quantities defined strictly within the
projected space spanned by the WFs. Our final expression for the total Berry
curvature, Eq.~(\ref{eq:bsum}), consists of three types of terms,
i.e., the $\overline{\Omega}$, $D$--$\overline{A}$, and $D$--$D$ terms.

We have applied this approach to calculate the AHC of bcc Fe.
While our  Wannier interpolation formalism, with its
decomposition (\ref{eq:bsum}),
is entirely independent of the choice of an all-electron or
pseudopotential method, we have chosen here a relativistic pseudopotential
approach that includes scalar relativistic effects as well as the 
spin orbit interaction.  We find that this
scheme successfully reproduces the fine details of the electronic
structure and of the Berry curvature in good agreement with a
previous calculation \cite{yao04} that used an all-electron LAPW
method.\cite{singh}  The computed AHC is also quite close to that
computed previously.\cite{yao04}

Interestingly, we found that more than 99\% of the total
Berry curvature is concentrated in the $D$--$D$ term of our formalism.
This term, given explicitly in Eq.~(\ref{eq:kubototm}), takes the form of 
a Kubo-like Berry curvature formula  for the ``tight-binding states.''
Thus we arrive at the very appealing result that
a Kubo picture defined within the ``exact tight-binding space'' gives an
excellent representation of the Berry curvature in the original 
{\it ab-initio} space. It is worth pointing out that, unlike the other
three terms, this term depends
exclusively on the Hamiltonian matrix elements between the Wannier orbitals,
and not on the position matrix elements. This result merits
further investigation, and may be relevant for
recent discussions in the
tight-binding literature on how to incorporate the coupling to
electromagnetic radiation in a tight-binding 
description.\cite{graf95,garm01,boykin01,foreman02}

Several directions for future studies suggest themselves.
For example, it would be desirable to obtain a better understanding
of how the AHC depends on the weak spin-orbit interaction.  As we
have seen, this weak interaction causes splittings and avoided
crossings that give rise to very large Berry curvatures in very
small regions of $k$-space.  There is a kind of paradox here.
Our numerical tests, as in Fig.~\ref{fig:lam}, demonstrate that
the AHC falls smoothly to zero as the spin-orbit strength $\lambda$
is turned off, suggesting that a perturbation theory in $\lambda$
should be applicable.  However, in the limit that $\lambda$
becomes small, the full calculation becomes {\it more difficult},
not less: the splittings occur in narrower and narrower regions
of $k$-space, energy denominators become smaller, and Berry
curvature contributions become larger (see Fig.~\ref{fig:bdbc}),
even if the {\it integrated} contribution is going to zero.
It would be of considerable interest, therefore, to explore ways
to reformulate the perturbation theory in $\lambda$ so that the
expansion coefficients can be computed in a robust and efficient
fashion.  Because the exchange splitting is much larger than the
spin-orbit splitting, it may also be of use to introduce
two separate couplings that control the strengths of the
spin-flip and spin-conserving parts of the spin-orbit
interaction respectively, and to work out the perturbation theory
in these two couplings independently.

Another promising direction is to explore whether the AHC can be
computed as a Fermi-surface integral using the formulation of
Haldane\cite{haldane04} in which an integration by parts is used
to convert the volume integral of the Berry curvature to a
Fermi-surface integral involving Berry curvatures or potentials.
Such an approach promises to be more efficient than the volume-integration
approach, provided that a method can be developed for carrying out an
appropriate sampling of the Fermi surface.  This is likely to be a
delicate problem, however, since the weak spin-orbit splitting causes
Fermi sheets to separate and reattach in a complex way at short
$k$-scales, and the
dominant contributions to the AHC are likely to come from precisely
these portions of the reconstructed Fermi surface that are the most
difficult to describe numerically.

In any case, even without such further developments, the present approach
is a powerful one.  It reduces the expense needed to do an extremely
fine sampling of Fermi-surface properties to the level where the
AHC of a material like bcc Fe can be computed on a workstation in a
few hours.  This opens the door to realistic calculations of the intrinsic
anomalous Hall conductivity of much more complex materials.
More generally, the techniques developed here for the AHE are readily
applicable to other problems in the physics of metals which also require
a very dense sampling of the Fermi surface or Brillouin zone. 
For example, an extension of these ideas
to the evaluation of the electron-phonon coupling matrix elements by
Wannier interpolation is currently under way.\cite{giustino}

\acknowledgments

This work was supported by NSF Grant DMR-0549198 and by
the Laboratory Directed Research and Development Program of Lawrence
Berkeley National Laboratory under the Department of Energy Contract
No. DE-AC02-05CH11231.

\appendix
\section{Alternative expression for the Berry curvature}
\label{app:alt-berry}

In this Appendix, we return to Eq.~(\ref{eq:om-a}) and rewrite it
in such a way that all of the large, rapidly varying contributions
arising from small energy denominators in the expression for
${D}_\alpha$, Eq.~(\ref{eq:ddef}), are segregated into a single term.
We do this by solving Eq.~(\ref{eq:Ata}) for ${D}_\alpha$
and substituting into Eq.~(\ref{eq:om-a}) to obtain
\begin{equation}
\Omega_{\alpha\beta}\ph =
\overline\Omega_{\alpha\beta}\ph -i \left[\overline A_\alpha\ph,\overline
   A_\beta\ph\right]
+i\left[A_\alpha\ph,A_\beta\ph\right]
\;.
\label{eq:om-b}
\end{equation}
Then only the last term will contain the large, rapid variations.
This equation could have been anticipated based on the fact that
the tensor
\begin{equation}
\widetilde{\Omega}_{\alpha\beta}= \Omega_{\alpha\beta}-i [A_\alpha,A_\beta]
\label{eq:gidef}
\end{equation}
is well known to be a gauge-covariant quantity;\cite{mead92,marzari97}
applying Eq.~(\ref{eq:utrans}) to $\widetilde{\Omega}_{\alpha\beta}$ then
leads directly to Eq.~(\ref{eq:om-b}).  

This formulation provides an alternative route to the
calculation of the matrix $\Omega_{\alpha\beta}\ph$:
evaluate $\widetilde{\Omega}\pw_{\alpha\beta}$ in the Wannier representation
using Eqs.~(\ref{eq:yya}-\ref{eq:yyb}) below,
convert it to $\widetilde{\Omega}\ph_{\alpha\beta}$ via Eq.~(\ref{eq:utrans}),
compute $A_\alpha\ph$ using Eq.~(\ref{eq:Ata}), and assemble
\begin{equation}
\Omega_{\alpha\beta}\ph=\widetilde{\Omega}_{\alpha\beta}\ph
   +i[A_\alpha\ph,A_\beta\ph]
\;.
\label{eq:alt}
\end{equation}
The large and rapid variations then appear only in the last term involving 
commutators of the $A$ matrices.

In Sec.~\ref{sec:bandsum}, we showed how to write the total Berry
curvature $\Omega_{\alpha\beta}(\k)$ as a sum over bands in such a way that 
potentially
troublesome contributions coming from small energy
denominators between pairs of occupied bands are explicitly
excluded, leading to Eq.~(\ref{eq:bsum}). The corresponding expression based
on Eq.~(\ref{eq:alt}) is
\begin{eqnarray}
\Omega_{\alpha\beta}(\k)&=&\sum_n f_n\,\widetilde{\Omega}_{nn,\alpha\beta}\ph
\nonumber\\
&&\quad  +\sum_{nm} (f_n-f_m)\,A_{nm,\alpha}\ph A_{mn,\beta }\ph
\;.
\label{eq:bsumalt}
\end{eqnarray}

Now, in addition to the four quantities given in
Eqs.~(\ref{eq:uu}-\ref{eq:xx}), we need a corresponding equation for
$\widetilde{\Omega}_{\alpha\beta}$.  After some manipulations, we find that
\begin{equation}
\widetilde{\Omega}_{nn,\alpha\beta}\pw(\k)=\sum_\R e^{i\k\cdot\R}\;
   w_{n,\alpha\beta}(\R)
\label{eq:yya}
\end{equation}
where
\begin{eqnarray}
w_{n,\alpha\beta}(\R) &=&
-i \sum_{\R'm}
   \langle{\bf 0}n|\hat{r}_\alpha|\R' m\rangle
   \langle\R' m|\hat{r}_\beta |\R n\rangle 
\nonumber\\ &&
+i \sum_{\R'm}
   \langle{\bf 0}n|\hat{r}_\beta |\R' m\rangle
   \langle\R' m|\hat{r}_\alpha|\R n\rangle 
\;.
\nonumber\\
\label{eq:yyb}
\end{eqnarray}
This formulation again requires the same basic ingredients as
before, namely, the Wannier matrix elements of $\hat{H}$ and
$\hat{r}_\alpha$.  In some respects it is a little more elegant
than the formulation of Eq.~(\ref{eq:bsum}). However,
the direct evaluation of $w_{n,\alpha\beta}$ in the Wannier
representation, as given in Eq.~(\ref{eq:yyb}), is not
as convenient because of the extra sum over intermediate WFs appearing
there; moreover, $w_{n,\alpha\beta}$ is longer-ranged than
the Hamiltonian and coordinate matrix elements.
Also, one appealing feature of the formulation of Section~\ref{sec:fdbc},
that more than 99\% of the effect can be recovered without using the
position-operator matrix elements, is lost in this reformulation.
We have therefore chosen to base our calculations and analysis on
Eq.~(\ref{eq:bsum}) instead.

It is informative to obtain Eq.~(\ref{eq:alt}) in a different way:
define the gauge-invariant band projection operator\cite{marzari97}
$\hat{P}_\k=\sum_{n=1}^M |u_{n\k}\rangle\langle u_{n\k}|$
and its complement $\hat{Q}_\k=1-\hat{P}_\k$. Inserting
$\hat 1=\hat{Q}_\k+\hat{P}_\k$ into Eq.~(\ref{eq:Owg}) in the
Hamiltonian gauge then yields directly Eq.~(\ref{eq:alt}) since, as can
be easily verified, Eq.~(\ref{eq:gidef}) may be written as
\begin{equation}
\widetilde{\Omega}_{nm,\alpha\beta}=
  i\langle \widetilde{\partial}_\alpha u_{n}|\widetilde{\partial}_\beta  u_{m}\rangle
 -i\langle \widetilde{\partial}_\beta  u_{n}|\widetilde{\partial}_\alpha
  u_{m}\rangle\;,
\label{eq:cov}
\end{equation}
where $\widetilde{\partial}_\alpha\equiv\hat{Q}\partial_\alpha$. The
gauge-covariance of $\widetilde{\Omega}_{\alpha\beta}$ follows directly from
the fact that $\widetilde\partial_\alpha$
is a gauge-covariant derivative, in the sense that
$|\widetilde{\partial}_\alpha u_n\ph\rangle=
\sum_{m=1}^M |\widetilde{\partial}_\alpha u_m\pw\rangle U_{mn}$
is the same transformation law as Eq.~(\ref{eq:twist}) for the Bloch
states themselves.
It is apparent from this derivation that as the number $M$ of WFs increases
and $\hat{P}_\k$ approaches $\hat 1$, the second term on the right-hand side
of Eq.~(\ref{eq:bsumalt}) increases at the expense of the first term.
Indeed, in the large-$M$ limit the entire Berry curvature is contained
in the second term. For the choice Wannier orbitals described in the
main text for bcc Fe, that term already accounts for 99.8\% of the
total AHC.

\section{Finite-difference approach}
\label{app:fda}

In this Appendix, we outline an alternative scheme for computing
the AHC by Wannier interpolation. The essential difference
relative to to the approaches described in Section~\ref{sec:fdbc} and
in Appendix~\ref{app:alt-berry} is that the needed $k$-space
derivatives are approximated
here by finite differences instead of being expressed analytically in
the Wannier representation.

This approach is most naturally applied to
the zero-temperature limit where there are exactly 
$N_\k$ occupied states at a given $\k$.
Instead of starting from the Berry curvature of each individual band
separately, as in Eq.~(\ref{eq:bcurv}), we find it convenient here to
work from the outset with the total Berry curvature
\begin{eqnarray}
        \Omega_{\alpha\beta}(\k)=\sum_{n=1}^{N_\k}\Omega_{nn,\alpha\beta}(\k)
\label{eq:omtot-a}
\end{eqnarray}
of the occupied manifold at $\k$
(the zero-temperature limit of Eq.~(\ref{eq:omtot})).
We now introduce a covariant derivative
$\widetilde\partial_\alpha^{(N_\k)}=\hat{Q}_\k^{(N_\k)}\partial_\alpha$
designed to act on the occupied states only; here
$\hat{Q}_\k^{(N_\k)}=\hat{1}-\hat{P}_\k^{(N_\k)}$ and
$\hat{P}_\k^{(N_\k)}=\sum_{n=1}^{N_\k}\,|u_{n\k}\rangle\langle u_{n\k}|$.
The only difference with respect to the definition of
$\widetilde{\partial}_\alpha$ in 
Appendix~A is that the projection operator here spans
the $N_\k$ occupied states only, instead of the $M$ states of the full
projected space. Accordingly, terms such as ``gauge-covariance'' and
``gauge-invariance'' are to be understood here in a restricted sense.
For example, the statement that $\widetilde\partial_\alpha^{(N_\k)}$ is a 
gauge-covariant derivative means that
under an $N_\k\times N_\k$ unitary rotation ${\cal U}(\k)$ between the
occupied states at $\k$ it obeys the transformation law
\begin{equation}
|\widetilde{\partial}^{(N_\k)}_\alpha u_{n\k}\rangle\rightarrow
\sum_{m=1}^{N_\k} |\widetilde{\partial}^{(N_\k)}_\alpha u_{m\k}\rangle\,
{\cal U}_{mn}(\k).
\end{equation}
(We will use calligraphic symbols to distinguish $N_\k\times N_\k$ matrices
such as $\cal U$ from their $M\times M$ counterparts such as $U$.)
We now define a gauge-covariant
curvature $\widetilde\Omega_{\alpha\beta}^{(N_\k)}(\k)$ by replacing
$\widetilde\partial$ by $\widetilde\partial^{(N_\k)}$ in Eq.~(\ref{eq:cov}).
Since the trace of a commutator vanishes, it follows from Eq.~(\ref{eq:gidef})
that Eq.~(\ref{eq:omtot-a}) can be written as
\begin{eqnarray}
\label{eq:partial_trace}
        \Omega_{\alpha\beta}(\k)={\rm Tr}^{(N_\k)}\,\left[\,\widetilde
        \Omega_{\alpha\beta}^{(N_\k)}(\k) \,\right],
\end{eqnarray}
where the symbol ${\rm Tr}^{(N_\k)}$ denotes the trace over the occupied 
states.

The advantage of this expression over Eq.~(\ref{eq:omtot-a})  is that the 
covariant derivative of a Bloch state can be approximated by
a very robust finite-differences formula:\cite{sai02,souza04}
\begin{eqnarray}
        \tilde \partial_{\k}^{(N_\k)}\rightarrow \sum_{\b}w_{b}{\b}\hat
        P_{\k,\b}^{(N_\k)}\; ,
\label{eq:dis}
\end{eqnarray}
where the sum is over shells of neighboring
$k$-points,\cite{marzari97}
as in Eq.~(\ref{eq:A_dis}),
and we have defined the gauge-invariant operator
\begin{eqnarray}
        \hat P_{\k,\b}^{(N_\k)}=\sum_{n=1}^{N_\k}|\widetilde
u_{n,\k+\b}\rangle \langle u_{n\k}|\;
\end{eqnarray}
in terms of the gauge-covariant ``dual states''
\begin{eqnarray}
                |\widetilde u_{n,\k,\b}\rangle=
                \sum_{m=1}^{N_\k}|u_{m,\k+\b}\rangle
\left (\cal Q_{\k+\b,\k}\right )_{mn}\;.
\end{eqnarray}
Here $\cal Q_{\k+\b,\k}$ is the inverse of the 
$N_\k\times N_\k$ overlap matrix,
\begin{eqnarray}
        {\cal Q}_{\k+\b,\k}=\left(\cal S_{\k,\k+\b}\right)^{-1}\;,
\end{eqnarray}
where
\begin{eqnarray}
        \left(\cal S_{\k,\k+\b}\right)_{nm}=\langle u_{n\k}
        | u_{m,\k+\b}\rangle\;.
        \label{eq:overlap}
\end{eqnarray}
The discretization (\ref{eq:dis}) is immune to arbitrary gauge phases
and unitary rotations among the occupied states. Because of that
property, the occurrence of band crossings and avoided crossings does
not pose any special problems.

Inserting Eqs.~(\ref{eq:dis}-\ref{eq:overlap}) into  
Eq.~(\ref{eq:partial_trace}) and using 
$\cal Q_{\k,\k+\b}=\cal Q^\dagger_{\k+\b,\k}$, we find that an
appropriate finite-difference expression for the total Berry
curvature is
\begin{equation}
\Omega_{\alpha \beta}^{(N_\k)}(\k)=
2\sum_{\b_1,\b_2}w_{b_1}\,w_{b_2}\,b_{1,\alpha}\,b_{2,\beta}
\,\Lambda_{\k,\b_1,\b_2},
        \label{eq:omega}
\end{equation}
where
\begin{equation}
\Lambda_{\k,\b_1,\b_2}=
       -{\rm Im}\,{\rm Tr}^{(N_\k)}\,
        \left[  {\cal Q}_{\k,\k+\b_1}{\cal S}_{\k+\b_1,\k+\b_2}
        {\cal Q}_{\k+\b_2,\k}\right]\,.
        \label{eq:Lambda}
\end{equation}
This expression is manifestly gauge-invariant, since both $\cal S$ and
$\cal Q$ are gauge-covariant matrices, i.e.,
${\cal S}_{\k,\k+\b}\rightarrow{\cal U}^\dagger(\k){\cal S}_{\k,\k+\b}
{\cal U}(\k+\b)$, and the same transformation law holds for
${\cal Q}_{\k,\k+\b}$.

Eqs.~(\ref{eq:omega}-\ref{eq:Lambda}) can be evaluated at an arbitrary point 
$\k$ once the overlap 
matrices ${\cal S}_{\k,\k+\b}$ are known. For that purpose we construct a 
uniform mesh of spacing $\Delta k$ in the immediate vicinity of $\k$, 
set up the needed shells of
neighboring $k$-points $\k+\b$ on that local mesh, and then evaluate
${\cal S}_{\k,\k+\b}$ by Wannier interpolation. Since the
WFs span the entire $M$-dimensional projected space, at this stage we revert
to the full $M\times M$ overlap matrices $S_{\k,\k+\b}$.
In the Wannier gauge they are given by a Fourier transform of the form
\begin{equation}
        \left({ S}^{\rm (W)}_{\k,\k+\b}\right)_{nm}=
\sum_{\R}e^{i\k \cdot \R}\langle {\bf 0} n |e^{i\b
\cdot (\R-\hat\r)}|\R m\rangle\;.
\end{equation}
For sufficiently small $\Delta k$, this can be approximated as
\begin{equation}
\left(S^{\rm (W)}_{\k,\k+\b}\right)_{nm}
\simeq\delta_{nm}-i\b\sum_{\R}e^{i\k \cdot \R}\langle {\bf 0} n |\hat\r|\R m\rangle\;.
\end{equation}
Note that the dependence of the last expression on $\Delta k$ is trivial,
since it only enter as a multiplicative prefactor. In practice one
chooses $\Delta k$ to be quite small, $\sim10^{-6}$\,a.u.$^{-1}$,
so as to reduce the error of the finite-differences expression.

In the Wannier gauge the occupied and empty states are mixed with one another,
because the WFs are partially occupied. In order to decouple the two subspaces
we perform the unitary transformation
\begin{eqnarray}
        {S}_{\k,\k+\b}^{\rm (H)}=U^{\dagger}(\k){S}_{\k,\k+\b}^{\rm (W)}
U({\k+\b})\;.
\label{eq:decouple}
\end{eqnarray}
This produces the full $M\times M$ overlap matrix in the Hamiltonian
gauge. The $N_\k\times N_\k$ submatrix in the upper left corner is
precisely the matrix ${\cal S}_{\k,\k+\b}\ph$ needed in Eq.~(\ref{eq:Lambda}).

Like the approach described in the main text, this approach
still only requires the WF matrix elements of the
four operators $\hat{H}$ and $\hat{r}_\alpha$ ($\alpha=x$, $y$, and $z$).
We have implemented it, and have checked that the
results agree closely with those obtained using
using the method of the main text. Although not as
elegant, this approach has the interesting feature
of circumventing the evaluation of the matrix $D_\alpha\ph$,
Eq.~(\ref{eq:ddef}). This may
be advantageous in certain special situations. For example, if a
parameter such as pressure is tuned in such a way that a $k$-space
Dirac monopole\cite{fang03} drifts to the Fermi surface,
the vanishing of the energy denominator in Eq.~(\ref{eq:ddef}) may result
in a numerical instability when trying to find the monopole
contribution to the AHC.

We conclude by noting that Eq.~(\ref{eq:Lambda}) is but one of many
possible finite-differences expressions, and may not even be the most 
convenient one to use
in practice. By recalling that the Berry curvature is the Berry phase
per unit area, one realizes that in the small-$\Delta k$ limit of
interest, the quantity $\Lambda_{\k,\b_1,\b_2}$ in Eq.~(\ref{eq:omega})
can be viewed as the discrete Berry phase $\phi$ accumulated along the 
small loop
$\k\rightarrow \k+\b_1\rightarrow\k+\b_2\rightarrow\k$. As is well-known, the
Berry phase around a discrete loop is defined as\cite{ksv93}
\begin{equation}
\phi=-\,{\rm Im}\,\ln\det
\left[
  {\cal S}_{\k,\k+\b_1}{\cal S}_{\k+\b_1,\k+\b_2}{\cal S}_{\k+\b_2,\k}
\right]\;.
\label{eq:dis_berry}
\end{equation}
It can be shown 
that $\phi=\Lambda_{\k,\b_1,\b_2}+
{\cal O}(\Delta k^2)$, so that for small loops the two formulas agree. 
Eq.~(\ref{eq:dis_berry}) has the practical advantage over
Eq.~(\ref{eq:Lambda}) that it does not require 
inverting the overlap matrix.



\begin{thebibliography}{10}
\bibitem{explan-kubo}

An advantage of the Kubo formula is that it can
easily be extended to obtain the frequency-dependence of the AHC.
\end{thebibliography}
\end{document}